\documentclass[acmsmall, nonacm]{acmart}
\usepackage[normalem]{ulem}
\usepackage{graphicx}
\usepackage{subfig}
\usepackage{multirow}
\usepackage{subfig}
\usepackage{textcomp}
\usepackage[export]{adjustbox}
\usepackage{multicol,lipsum}
\usepackage{xurl}
\usepackage{xcolor}
\usepackage[linesnumbered]{algorithm2e}
\DontPrintSemicolon 
\AtBeginDocument{%
  \providecommand\BibTeX{{%
    \normalfont B\kern-0.5em{\scshape i\kern-0.25em b}\kern-0.8em\TeX}}}

\makeatletter
\let\@authorsaddresses\@empty
\makeatother

\begin{document}

%%
%% The "title" command has an optional parameter,
%% allowing the author to define a "short title" to be used in page headers.
%\title{Model-driven Latency-sensitive AI Inference on Shared Edge Accelerators}
\title{Model-driven Cluster Resource Management for AI Workloads in Edge Clouds}

%%
%% The "author" command and its associated commands are used to define
%% the authors and their affiliations.
%% Of note is the shared affiliation of the first two authors, and the
%% "authornote" and "authornotemark" commands
%% used to denote shared contribution to the research.

%\author{Anonymous Author(s)}
\author{Qianlin Liang}
\email{qliang@cs.umass.edu}
\affiliation{%
  \institution{University of Massachusetts Amherst}
  \city{Amherst}
  \state{MA}
  \country{USA}
  \postcode{01002}
}
%\orcid{1234-5678-9012}

\author{Walid A. Hanafy}
\email{whanafy@cs.umass.edu}
\affiliation{%
  \institution{University of Massachusetts Amherst}
%  \streetaddress{P.O. Box 1212}
  \city{Amherst}
  \state{MA}
  \country{USA}
  \postcode{01002}
}

\author{Ahmed Ali-Eldin}
\email{ahmeda@cs.umass.edu}
\affiliation{%
  \institution{University of Massachusetts Amherst}
%  \streetaddress{P.O. Box 1212}
  \city{Amherst}
  \state{MA}
  \country{USA}
  \postcode{01002}
}

\author{Prashant Shenoy}
\email{shenoy@cs.umass.edu}
\affiliation{%
  \institution{University of Massachusetts Amherst}
%  \streetaddress{P.O. Box 1212}
  \city{Amherst}
  \state{MA}
  \country{USA}
  \postcode{01002}
}

%%
%% By default, the full list of authors will be used in the page
%% headers. Often, this list is too long, and will overlap
%% other information printed in the page headers. This command allows
%% the author to define a more concise list
%% of authors' names for this purpose.
%\renewcommand{\shortauthors}{Trovato and Tobin, et al.}

%%
%% The abstract is a short summary of the work to be presented in the
%% article.
\begin{abstract}
	%Machine Learning inference is central to many edge computing applications, including, autonomous vehicles, AR, and VR. These applications have tight performance and latency requirements and hence depend on the low-latency to the edge when offloading computations. Multiple novel edge accelerators such as Nvidia' Jetson Nanos, and Google's Tensor Processing Units (TPU), have been developed to support these workloads. However, the performance characteristics of these novel devices are not well understood. Such an understanding is essential for managing workloads and resources at the edge, and building efficient schedulers. In this paper, we develop closed-form queuing models to capture the performance of three such devices, namely, Jetson Nanos, TPUs, and small-sized GPUs suitable for edge deployments. We validate the models using a wide-range of pre-trained popular inference Deep Neural Network models.  In addition, we discuss the performance limitations of each of these devices. We finally show how our queuing models can be used for performance-aware DNN model placement, migration, and co-location.
	Since emerging edge applications such as Internet of Things (IoT) analytics and augmented reality have tight latency constraints, hardware AI accelerators have been recently proposed to speed up deep neural network (DNN) inference run by these applications.  Resource-constrained  edge servers and accelerators tend to be multiplexed across multiple IoT applications, introducing the potential for performance interference between latency-sensitive workloads. In this paper, we design analytic models to capture the performance of DNN inference workloads on shared edge accelerators, such as GPU and edgeTPU, under different multiplexing and concurrency behaviors.  After validating our models using extensive experiments, we use them to design various cluster resource management algorithms to intelligently manage multiple applications on edge accelerators while respecting their latency constraints. We implement a prototype of our system in Kubernetes and  show that our system can host 2.3X more DNN applications in heterogeneous multi-tenant edge clusters with no latency violations when compared to traditional knapsack hosting algorithms.
\end{abstract}

%%
%% The code below is generated by the tool at http://dl.acm.org/ccs.cfm.
%% Please copy and paste the code instead of the example below.
%%
%\begin{CCSXML}
%<ccs2012>
% <concept>
%  <concept_id>10010520.10010553.10010562</concept_id>
%  <concept_desc>Computer systems organization~Embedded systems</concept_desc>
%  <concept_significance>500</concept_significance>
% </concept>
% <concept>
%  <concept_id>10010520.10010575.10010755</concept_id>
%  <concept_desc>Computer systems organization~Redundancy</concept_desc>
%  <concept_significance>300</concept_significance>
% </concept>
% <concept>
%  <concept_id>10010520.10010553.10010554</concept_id>
%  <concept_desc>Computer systems organization~Robotics</concept_desc>
%  <concept_significance>100</concept_significance>
% </concept>
% <concept>
%  <concept_id>10003033.10003083.10003095</concept_id>
%  <concept_desc>Networks~Network reliability</concept_desc>
%  <concept_significance>100</concept_significance>
% </concept>
%</ccs2012>
%\end{CCSXML}
%
%\ccsdesc[500]{Computer systems organization~Embedded systems}
%\ccsdesc[300]{Computer systems organization~Redundancy}
%\ccsdesc{Computer systems organization~Robotics}
%\ccsdesc[100]{Networks~Network reliability}

%%
%% Keywords. The author(s) should pick words that accurately describe
%% the work being presented. Separate the keywords with commas.
%\keywords{datasets, neural networks, gaze detection, text tagging}

%%
%% This command processes the author and affiliation and title
%% information and builds the first part of the formatted document.
\maketitle

% Introduction
\section{Introduction}
\label{sec:intro}

Recent technological advances have resulted in the emergence of new applications such as mobile Augmented Reality (mobile AR)~\cite{chen2018marvel} and  Internet of Things (IoT) analytics~\cite{salehe2019videopipe}. A common characteristic of these applications is that their data needs to be processed with tight latency constraints. 
%The traditional technique of cloud offloading, where data is sent to cloud servers for processing, incurs WAN delays and may not meet these tight latency-sensitive requirements.
%Processing this data in real-time, and close to the point of generation, can reduce network latency and bandwidth usage on WAN links. 
Consequently, edge computing, where edge resources can process this data  close to the point of generation and at low latencies, has emerged as a popular approach for meeting the needs of these emerging applications \cite{satyanarayanan2009case}. 

It is increasingly common for such IoT applications to use AI inference as part of their data processing tasks. In contrast to \emph{model training}, which involves training a machine learning (ML) model in the cloud, typically using large GPUs, \emph{model inference} involves executing a previously trained  model for inference (i.e., predictions) over new data.  For instance, video or audio data generated by IoT devices such as AR headsets, smart cameras, or smart speakers can be sent to a trained machine learning model for inference tasks such as object or speech recognition.  The model, which is often a deep neural network (DNN), runs on an edge server to provide low-latency inference processing to the application.
%\textcolor{blue}{Although AI \emph{inference} is less computationally intensive than model \emph{training}, it is also amenable to the use of special-purpose hardware to accelerate inference tasks.}

Edge clouds, which extend cloud computing to the edge, are an increasingly popular approach for running low-latency inference for IoT applications. Edge clouds consist of small edge clusters that are deployed at a number of edge locations, where each edge cluster hosts multiple IoT applications. To efficiently run deep learning inference in constrained edge environments, such edge clouds have begun to use accelerators that are capable of executing DNN models using specialized hardware. 
Examples of  edge accelerators include Google's EdgeTPU \cite{TPUPatent}, Nvidia Jetson line of embedded GPUs \cite{nvidia-jetson}, and Intel's Movidius Vision Processing Units (VPUs) \cite{intel-vpu}. When equipped with such hardware accelerators, edge cloud servers can significantly improve DNN inference tasks' latency--- similar to how cloud servers use larger GPUs to speed up DNN training. 
As AI inference for IoT data processing gains popularity, the use of such accelerators to optimize DNN inference  is likely to become commonplace in edge cloud environments.

Executing AI inference over IoT data in  edge clouds raises new challenges. Similar to traditional cloud platforms, edge clouds will also be multi-tenant in nature, which means that each edge cloud server will run multiple tenant applications. These applications share the hardware resources of edge servers, including  accelerators \cite{gslice, gpushare, olympian}. While conventional resources such as CPU and even server GPUs~\cite{gpu-virtualization} support virtualization features to enable them to be multiplexed across applications, edge accelerators lack such hardware features. Consequently, multiplexing a DNN accelerator across multiple tenant applications can result in performance interference due to the lack of isolation mechanisms such as virtualization, which  can degrade the response times seen by latency-sensitive IoT applications.
This motivates the need for developing new cluster resources management techniques for  efficiently multiplexing shared edge cloud resources across latency-sensitive applications.

To address the aforementioned challenges,  in this paper, we present \emph{Ibis}, a model-driven cluster resource management system for edge clouds. Ibis is designed to multiplex  cluster resources, such as DNN accelerators, across multiple edge applications while limiting the performance interference between co-located tenants.   Ibis uses a principled resource management approach based on analytic queueing models of hardware accelerators, such as edgeTPUs and edge GPUs, that are extensively experimentally validated on real edge clusters. These models are incorporated into Ibis' cluster resource manager and used to manage the online placement and dynamic migration of edge applications. In designing, implementing, and evaluating Ibis, our paper make the following research contributions. 

%we present model-driven approaches for efficient sharing of edge servers and accelerators across multiple latency-sensitive DNN-based IoT applications.  Our paper makes the following contributions.

First, we develop analytical models, based on queueing theory, to estimate the response times seen by co-located DNN-based IoT applications running on shared edge servers with accelerators.  We develop models to capture a range of multiplexing behaviors such as  FCFS, processor sharing, batching, and multi-core parallelism that are seen when sharing edge GPUs and TPUs. We experimentally validate our models using two dozen different DNN models, drawn from popular  model families such as AlexNet, ResNet, EfficientNet, Yolo, Inception, SSG, VGG, and DenseNet, and a variety of IoT workloads. Our extensive validation demonstrates the abilities of our models to capture performance interference and accurately predict response times of co-located applications.

%Second, drawing inspiration from multi-tier web applications, we present a two-tier architecture for DNN inference applications, using OS containers, that limit the performance interference from co-located tenants while enabling each tenant to run arbitrary application code on edge servers.
Second, we present an edge cluster resource manager that uses our analytic models for resource management tasks such as online placement and dynamic migration. Unlike traditional placement problems, which can be viewed as an online knapsack, we formulate a new problem called online knapsack with latency constraints for  DNN application placement onto shared accelerators results. We present greedy heuristics that use our analytic models for latency-aware placement  and migration in heterogeneous clusters.

%we use our analytic models to develop resource management algorithms for edge accelerator clusters performing tasks such as DNN model placement and co-location  by choosing an appropriate accelerator type. These management decisions are based on the DNN characteristics, the most-suitable accelerator node for the DNN, and the  utilization,   so as to meet the latency  requirements of each applications and limit performance interference.

Third, we implement a prototype of Ibis on a Kubernetes-based edge cluster and conduct detailed experiments to demonstrate the efficacy of our model-driven cluster resource management approach. Our results show that Ibis can host up to 2.3$\times$ the number of DNN models in heterogeneous edge clusters when compared to traditional knapsack hosting algorithms, and can dynamically mitigate hotspots using edge migration. 

\section{Background}
\label{sec:background}

In this section, we provide background on edge computing and edge accelerators.  % and AI workloads on edge servers.

\subsection{Edge Computing and AI Inference for IoT}
Edge clouds are a form of edge computing that involves deploying computing and storage resources at the edge of the network to provide low latency access to users \cite{Satyanarayanan2017}. 
% \textcolor{red}{Ahmed: I suggest to cut all the blue text. }
%In contrast to traditional cloud computing that are deployed large data centers, edge clouds  involve running smaller server clusters at the edge of the network \cite{Satyanarayanan2017}. 
%Edge clouds seek to provide the benefits of a traditional cloud in the context of edge computing where end customers do not need to deploy their own edge infrastructure and can instead  rent edge resources from the edge cloud provider when needed and run their edge applications on these edge servers.
%Relocated
However, each edge cluster is smaller, and hence more resource-constrained than traditional cloud data centers. Analogous to cloud platforms that run multiple tenant applications in a server cluster, each edge cluster and edge server, is multiplexed across multiple applications---to maximize the utilization of scarce edge resources.  Since many edge applications  are inherently latency-sensitive, it is important to limit performance interference between co-located tenant applications. This can be achieved through the use of resource isolation mechanisms (e.g., virtualization), where available, and by carefully limiting the utilization and sharing of each server across tenant applications.
%Our work uses analytic model-driven resource management algorithms to characterize the performance interference between multi-tenant edge applications  and meet their service level objectives (SLOs) on latency.  
\footnote{Note that the term ``model'' in this paper refers to both a deep neural network (DNN) inference model as well as an analytic queueing model; the two are distinct and we disambiguate them in the text as DNN inference and analytic/queueing model.}

Our work focuses on emerging edge IoT applications, such as mobile AR, visual analytics over live videos from smart cameras, and voice assistants (e.g., Alexa, Siri) that run on smart speakers.  Since these applications interact with users, it is necessary to process IoT data at low latencies to provide high user responsiveness, which imposes latency constraints on edge processing. A common characteristic of many IoT applications is that they employ machine learning models, often in the form of a Deep Neural Network (DNN),  to process their data at the edge.
% \textcolor{red}{Ahmed: Suggest to cut the blue text below. Repetition from the intro.}
% \textcolor{blue}{A common characteristic of applications such as mobile AR, video analytics, and speech recognition is that they employ machine learning models, often in the form of a Deep Neural Network (DNN),  for their data processing tasks. Typically, such a model is first trained on cloud servers using a large amount of training data \cite{Hafeez2020}, and the trained model is then deployed on edge servers for performing inference over data from mobile or IoT devices.}
Recent advances in computer vision technology have yielded a number of sophisticated and highly effective DNNs for common image processing tasks such as classification, object detection, and segmentation \cite{vision-textbook}. Advances in the field have allowed practitioners to provide a library of pre-trained DNNs such as ResNet \cite{resnet}, Inception \cite{inception}, MobileNet \cite{MobileNet}, and Yolo \cite{yolo}, among others. An edge cloud developer can simply  use one of these pre-trained DNN models  within their application for performing inference  tasks such as object detection or recognition over images. Alternatively, the developer can train a custom  DNN for their application using easy-to-use ML frameworks such as TensorFlow \cite{tensorflow}, PyTorch \cite{pytorch}, and Caffe \cite{caffe} and   datasets such as ImageNet \cite{imagenet} and CIFAR \cite{cifar}.

% has made it feasible for users to train their own custom DNN models and deploy them for inference tasks. 
%Advances in speech signal processing and natural language processing have yielded similar models for audio data \cite{audio-models}.
%These advances have led to the emergence of new class of edge workload that involves \emph{real-time inference} using a  trained DNN model at the edge in contrast to cloud inference, which is also becoming commonplace).
%As noted earlier, applications such as mobile AR, and video/audio analytics are latency-sensitive and bandwidth-intensive, making them well suited for edge computing, yielding edge inference workloads (in contrast to cloud inference, which is also becoming commonplace).

\subsection{Edge Inference Accelerators}
% \textcolor{red}{Ahmed: This text is again mostly a repetition from above. Suggest to minimize to `` We next discuss some of the most common edge accelerators available today.''}
% \textcolor{blue}{The growing popularity of edge inference workload has led to the design of special-purpose hardware specifically for inference workloads. Similar to how GPUs have long been used in cloud clusters to accelerate the training of ML models, these hardware accelerators seek to  accelerate DNN model inference at execution time. Many such edge inference accelerators are available today, including Nvidia's Jetson edge GPUs \cite{nvidia-jetson}, Google's edge TPUs \cite{Cass2019TakingAT} and Intel's  Movidius VPUs. We focus on two of the more popular edge inference accelerators here, namely edge GPUs and edge TPUs.}
The growing popularity of DNN inference at the edge has led to the design of special-purpose accelerator hardware such as Nvidia's Jetson GPUs \cite{nvidia-jetson} and Google's edgeTPUs \cite{TPUPatent}. Edge cloud servers are beginning to employ such hardware for efficient inference execution.  We provide a brief overview of these devices here.  %\cite{Amert2017, Yang2018} 
%and refer the reader to \cite{Amert2017, Yang2018} for a detailed discussion. 

\subsubsection{Edge GPUs}
\begin{figure*}
	\centering
    \captionsetup[subfigure]{position=b}
    \subfloat[Jetson Nano GPU]{
	    \centering
		\includegraphics[width=0.3\textwidth, valign=c]{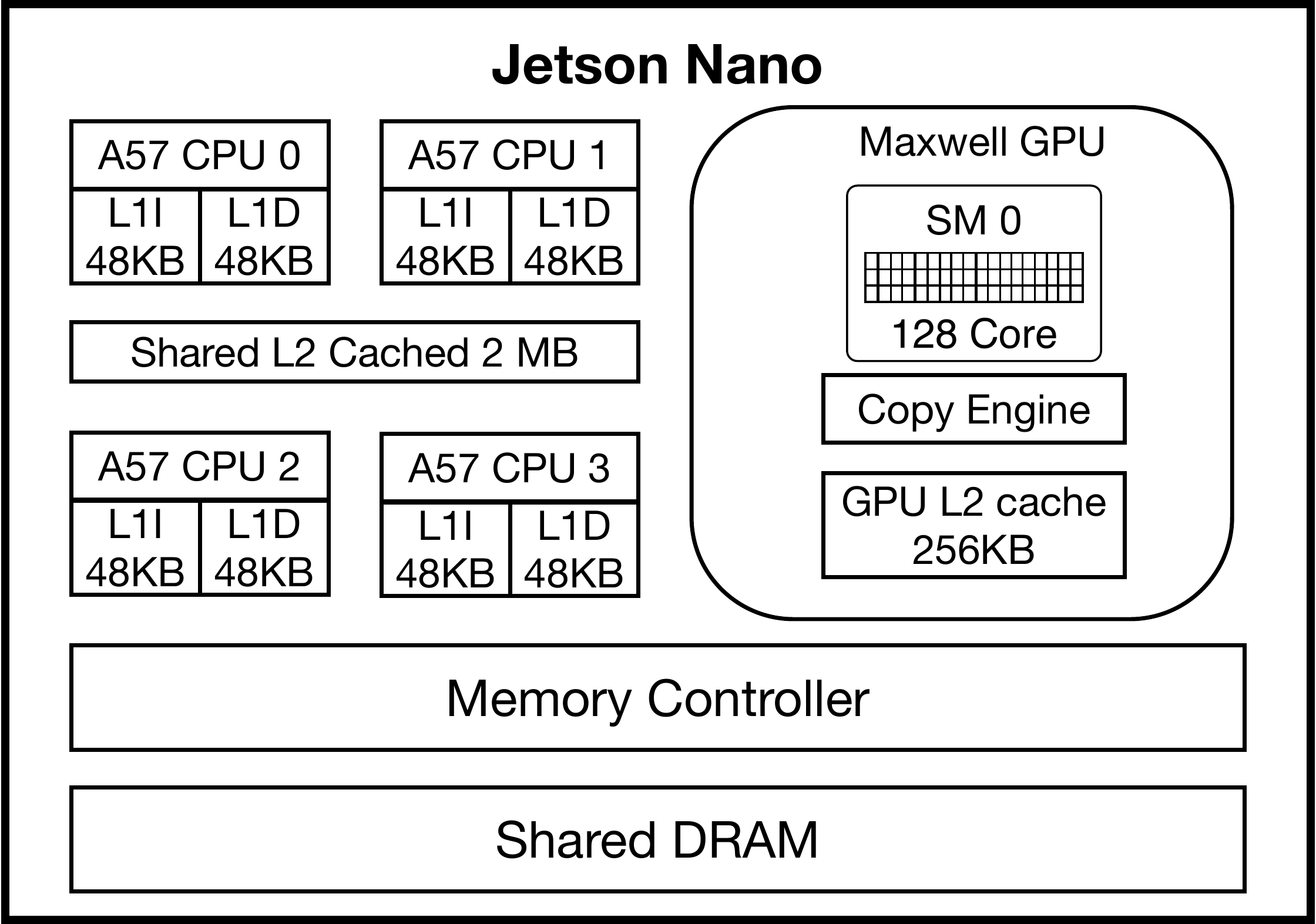}\label{fig:nano-arch}
	}%
	%\hspace
	\subfloat[GeForce GTX 1080 GPU]{
	    \centering
		\includegraphics[width=0.36\textwidth, valign=c]{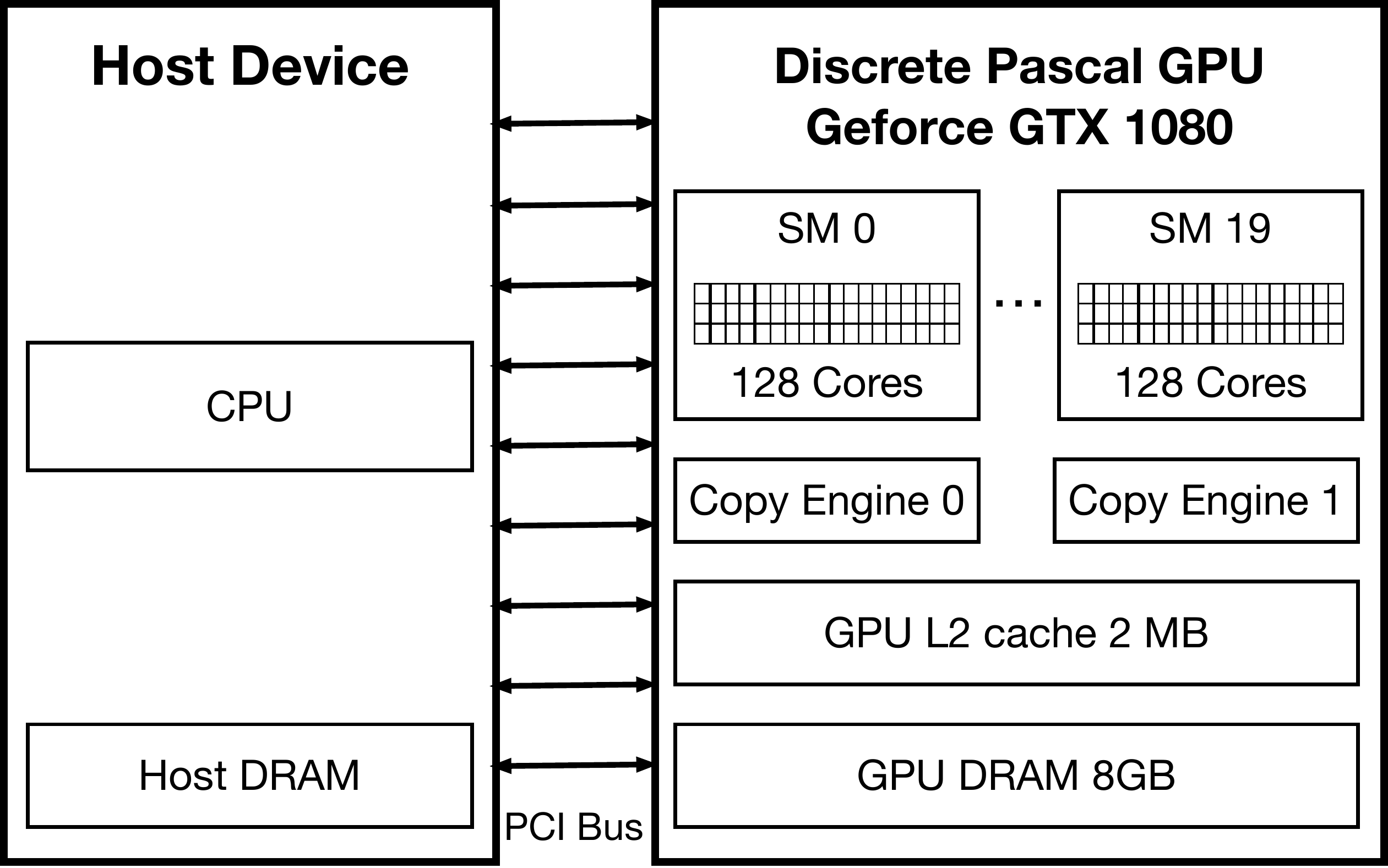}\label{fig:gtx1080-arch}
	}%
	%\hspace
	\subfloat[Google's edgeTPU]{
	    \centering
		\includegraphics[width=0.3\textwidth, valign=c]{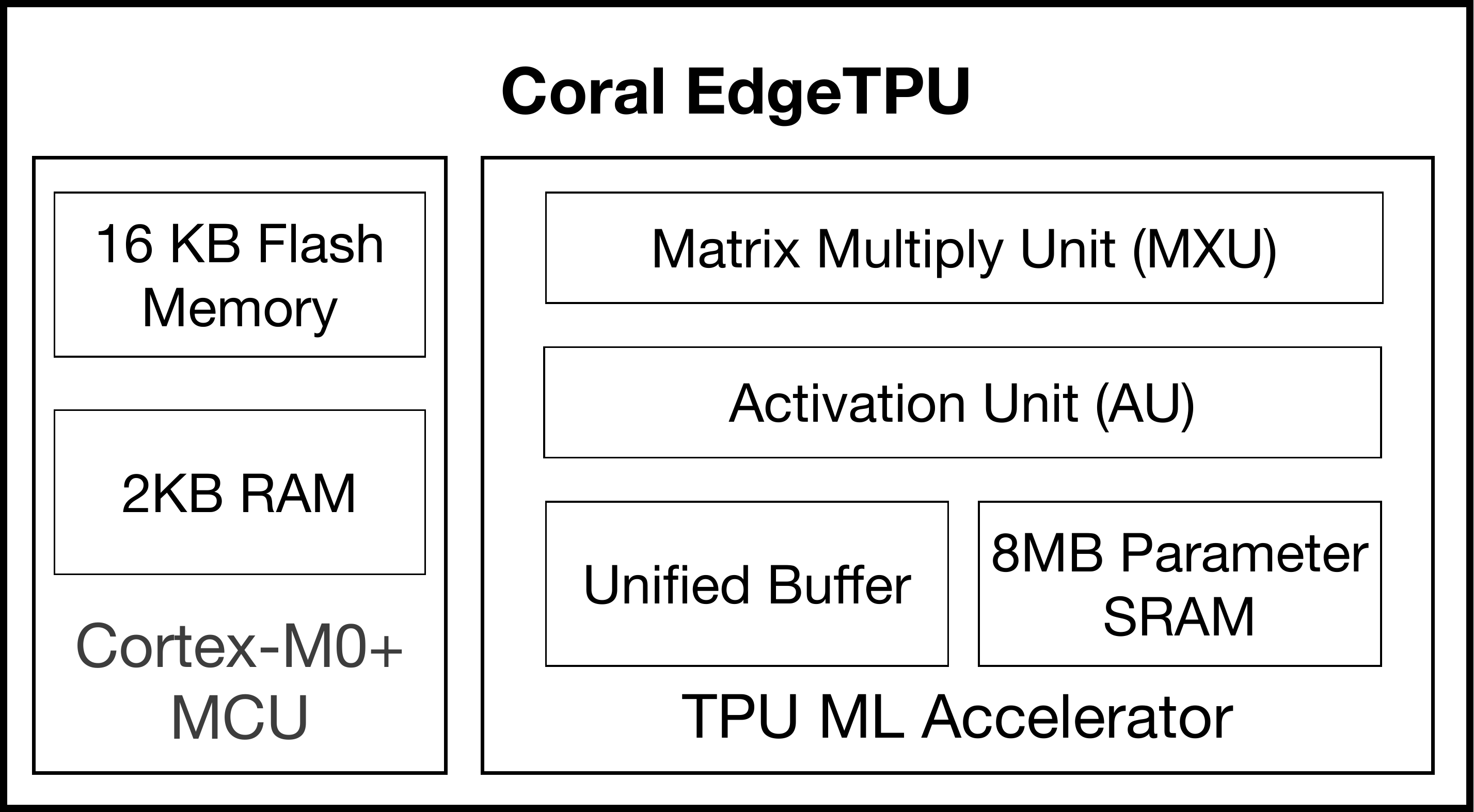}\label{fig:edgetpu_arch}
	}
    \captionsetup{belowskip=0pt, aboveskip=0pt}
	\caption{Architectural depiction of edge GPU, discrete GPU, and edgeTPU. }

\end{figure*}

Today's GPUs come in three main flavors, each targeting a different application workload. Server GPUs are high-end GPUs that are designed to accelerate parallel scientific computations or speedup the training of machine learning models. Discrete GPUs are designed for gaming as well as scientific desktop applications (e.g., CAD) and have also been recently used for edge processing~\cite{zhang2017towards,lin2016cloudfog} due to their interesting multiplexing capabilities. Finally, embedded GPUs are designed for edge (or on-device) applications. They have a low power footprint and are less capable than server or discrete GPUs, but well-suited for edge processing workloads. For example, Nvidia's Jetson family \cite{nvidia-jetson} of embedded GPUs are designed for running AI inference  at the edge. 

Figure \ref{fig:nano-arch} depicts the architecture of a Jetson Nano edge GPU---the smallest embedded GPU in the Jetson family. All GPUs in the Jetson family are \textit{integrated} GPUs that integrate the CPU, GPU, and memory onto a single System-on-Chip (SoC). The figure shows that the device has 4 ARM-based CPU cores and a GPU comprising one Streaming Multiprocessor (SM) with 128 CUDA (GPU) cores. Notably, the device has 4 GBs of RAM that is shared between the CPU and the GPU. Like any Nvidia GPU, the Jetson Nano runs programs written in CUDA \cite{cuda}. 

%A CUDA program such as DNN model inference runs as a normal OS process on the CPU and invokes GPU processing for accelerating data-parallel tasks.

%Each CPU process is associated with a CUDA \emph{context} \cite{cuda}. The process executes or offloads various sub-tasks to the GPU during its overall execution, each of which can be viewed as a small CUDA program and is referred to as a \emph{kernel}. Kernels are submitted to \emph{stream} for execution. Kernels within a stream are executed sequentially; however, each kernel uses data parallelism to exploit a large number of cores on the GPU.

Each CPU process is associated with a CUDA \emph{context} that is responsible for offloading  compute-intensive functions, referred to as \emph{kernels}, to the GPU.  Kernels are submitted as a \emph{stream} and executed sequentially. Nvidia GPUs support two basic types of concurrency, namely multi-processing (MP) and multi-threading (MT) \cite{Amert2017, Yang2018}. In multiprocessing, the GPU is time-shared between processes---processes (CUDA contexts) take turns to execute on the GPU for a time slice. On the other hand, in multi-threading, multiple threads, each associated with a separate CUDA context, can execute  kernels concurrently via thread-level parallelism. 

%Nvidia GPUs support two basic types of concurrency, namely multiprocessing (MP) and multi-threading (MT) \cite{Amert2017, Yang2018}. In the multiprocessing case, multiple OS processes, each associated with an independent CUDA context, can concurrently issue requests (in the form of kernels) to the GPU. The GPU uses time sharing to execute kernels from independent processes (while sequentially processing kernels requests issued by a single process) \cite{MPS}. While time-sharing enables concurrent processing of kernels from independent processes, it should be noted that only a single CUDA context can execute on the GPU at any time for a time slice and uses data parallelism to utilize GPU cores. 

%On the other hand, in the multi-threading case, multiple threads within a single OS process can issue kernel requests to the GPU. In this case, the process creates multiple CUDA contexts and associates them with its threads, which allows threads-level parallelism to issue kernel requests in parallel via multiple CUDA contexts.

Researchers have noted that multi-threaded execution introduces significant synchronization overheads on GPUs, with increased blocking and non-determinism for latency-sensitive tasks \cite{Amert2017, Yang2018}. Since such non-deterministic blocking behavior is problematic for latency-sensitive edge applications, we focus our work on process-level concurrency, which has been shown to provide more predictable behavior \cite{Amert2017, Yang2018}. In our case, this implies a separate process executes each DNN model, and different processes can issue concurrent inference requests execute using time-sharing on GPU cores.

\subsubsection{Discrete GPUs}
 
Discrete GPUs support additional multiplexing capabilities that are useful in multi-tenant edge settings~\cite{zhang2017towards,lin2016cloudfog}.
Discrete GPUs, such as GeForce 1080, are higher-end GPUs with their own on-GPU memory that is separate from CPU's RAM (see Fig \ref{fig:gtx1080-arch}).
As shown, the 1080 GPU has 20 SMs, 2560 GPU cores, and 8GM RAM, which is significantly greater than Jetson GPUs. Discrete GPUs support both multi-processing and multi-threading like their embedded counterparts. In addition, they also support Nvidia's Multi-Process Service (MPS) that enables true parallelism across concurrent GPU requests from independent processes \cite{MPS}. When MPS is enabled in a GPU, all kernel requests are forwarded to MPS for scheduling. The MPS system partitions the GPU cores and memory across CUDA contexts and schedules kernels for execution on each partition in parallel. In our case, this implies that DNN inference of two DNN models can execute in parallel the increasing utilization of the GPU cores (while in embedded GPUs, they execute using time-sharing but not in parallel). 

\subsubsection{EdgeTPU}\label{sec:edge_tpu_device}
EdgeTPU is an ASIC designed by Google for high performance DNN inference  using very low power. In contrast to GPUs, which are optimized for performing floating-point operations,  EdgeTPU uses 8-bit integers for computation and requires the DNN models to be quantized to 8-bit \cite{TPUPatent}. Employing quantization greatly reduces the hardware footprint and energy consumption of the EdgeTPU. 

Figure \ref{fig:edgetpu_arch} shows the architecture of EdgeTPU \cite{TPUPatent}. In contrast to CUDA cores, \textit{Matrix Multiply Unit} (MXU) is the heart of EdgeTPU. It employs a systolic array architecture, which reuses inputs many times without storing them back to a register. By reducing access to registers, MXU is optimized for power and area efficiency for performing matrix multiplications and allows EdgeTPU to perform 4 trillion fixed-point operations per second (4 TOPS) using only 2 watts of power.  EdgeTPU is designed specifically for DNN inference and is less general than GPUs. Its design is strictly deterministic and has a much smaller control logic than GPU. Thus, it can only run one task at a time in a non-preemptive FCFS manner.  Unlike GPUs, time-sharing and multiprocessing are not supported.

\section{Analytic Models for Inference Workloads} 
\label{sec:queue-models}
% Please add the following required packages to your document preamble:
% \usepackage{graphicx}
\begin{table}[b]%
\centering
\resizebox{0.6\textwidth}{!}{%
\begin{tabular}{|c|c|c|}
\hline
\textit{\textbf{Symbol}} & \textit{\textbf{Description}}          & \textit{\textbf{Notes}}                         \\ \hline
$\lambda$                & Arrival Rate                           & $\lambda = \sum \lambda_i$                      \\ \hline
$\mu$                    & Service Rate                           & $\mu = 1/S$                                     \\ \hline
$S$                      & Service time                           & $S = \sum \lambda_i S_i / \sum \lambda_i$ \\ \hline
$c$                      & number of cores                        &                                                 \\ \hline
$\rho$                   & Utilization                            & $\rho = \lambda/c\mu$                                  \\ \hline
$E[w]$                   & Expected Waiting Time                           &                                                 \\ \hline
$E[R]$                   & Expected Response Time                          & $E[w] + S$                                      \\ \hline
$P(M_i)$                 & Probability of a request for DNN $i$ & $P(M_i) = \lambda_i/\lambda$                    \\ \hline
$e_i$                    & Execution Time for DNN $i$           &                                                 \\ \hline
$o_i$                    & Context Switch overhead for DNN $i$          &                                                 \\ \hline
\end{tabular}%
}
\caption{Used Notations}
\captionsetup{belowskip=-30pt, aboveskip=-30pt}
\label{tab:symbols}
\end{table}

The goal of our work is to design a cluster resource manager for edge clouds that can efficiently multiplex edge server resources, and specifically accelerator resources, across multiple applications. Our system employs a \emph{model-driven resource management} approach, where we first design analytic models of edge workloads and then use these models to design practical cluster resource management algorithms. In this section, we design analytic models based on elementary queueing theory and use extensive experimentation to show that queueing   models can (i) capture a range of multiplexing behavior seen in real-world accelerators such as edge GPUs and edgeTPUs, and (ii) accurately estimate the response times of a broad range of real-world DNN models, and (iii) can capture the impact of interference from co-located applications on application response times.
%This section presents analytic models  to capture the response times seen by two-tiered DNN inference applications.  
For readability,   Table \ref{tab:symbols} summarizes the notation  and common equations  used in our models.
% In this section, we present queuing models that capture the behavior of our two-tier container architecture running on different types of edge nodes with different hardware accelerators. Our model assumes that the edge node receives an inference request (e.g., an image or a batch of images) over the network. The data goes through pre-processing on the CPU and then subjected to inference processing on GPU/TPU. Given our two-tier architecture, we assume that the bulk of the CPU load (pre-processing) occurs in the frontend container, and all of the inference processing occurs in the backend container.

% In the rest of this section, we describe how to model the CPU and inference processing using queuing theory.

\subsection{Network of Queues Model}

Queueing theory has been used to analytically model the behavior of web applications \cite{Urgaonkar2005, cloudscale, Grandhi2019Queueing}, server farms\cite{HarcholBalter2005MultiServerQS, serverFarmsHarchol} and cloud computing \cite{waiting-game}. Here we use it to analytically model concurrent DNN applications running on edge servers with accelerators such as GPU and TPU. 

To do so, we assume each edge server has at least one accelerator and runs $k$ concurrent DNN applications, $k \geq 1$. We assume that application $i$ receives DNN inference requests at rate $\lambda_i$ from an IoT device and specifies a mean response time  $R_i$ that should be provided by the edge cloud. We assume that each inference request undergoes a combination of CPU and GPU/TPU processing by the application. 

\begin{figure*}[t]
	\centering
	%\subfloat[]{
%		%\includegraphics[width=0.25\textwidth]{imgs/inference_as_a_service.pdf}\label{fig:backend_agg}
		%\includegraphics[width=0.5\textwidth]{imgs/backend_agg.pdf}\label{fig:two_tier_archiecture}
	%}\hspace*{0.01in}
	\subfloat[]{
	\includegraphics[width=0.21\textwidth]{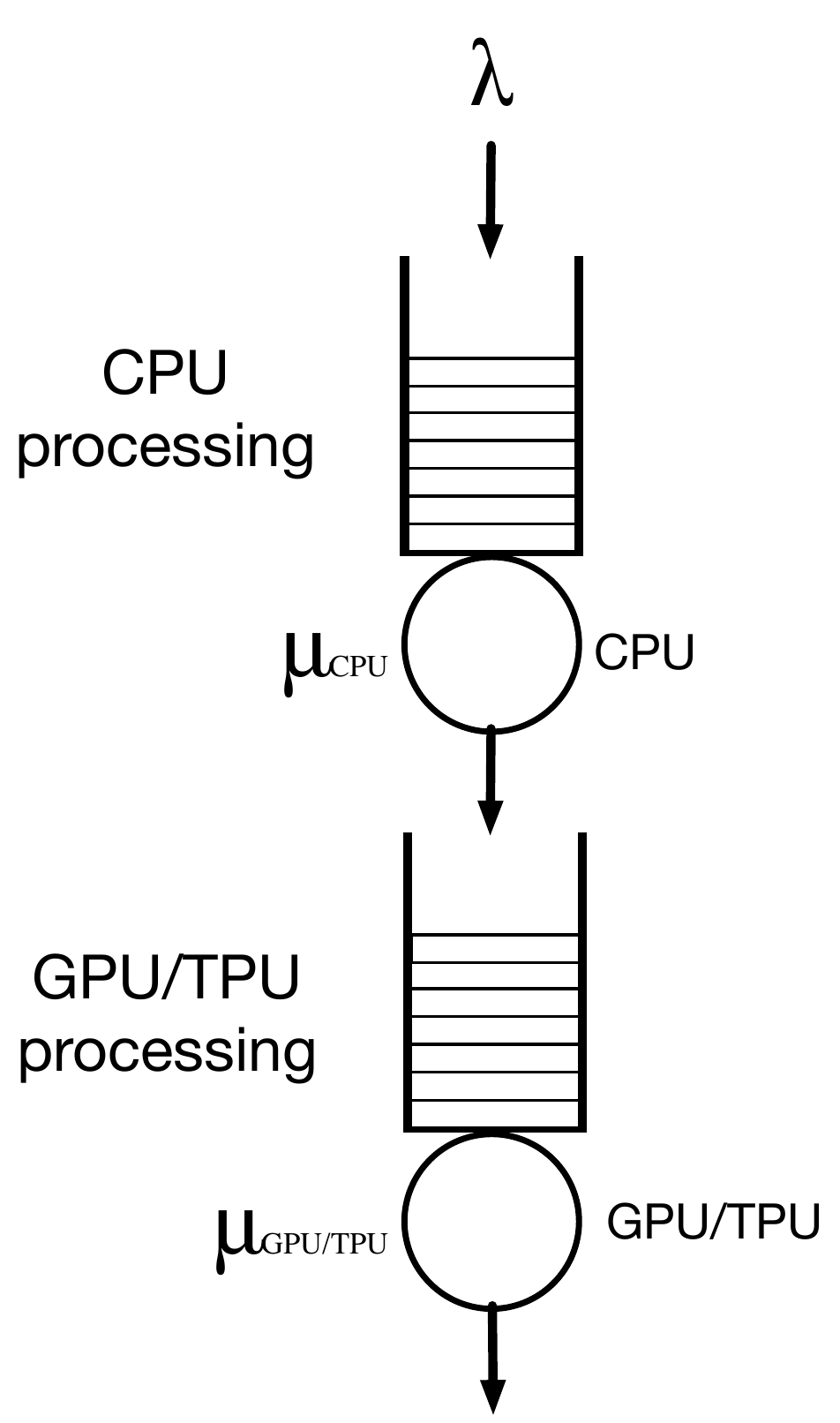}
	\label{fig:tandem_queue_arch}
	}%
\hspace{0.5in}
	\subfloat[]{
		\includegraphics[width=0.432\textwidth]{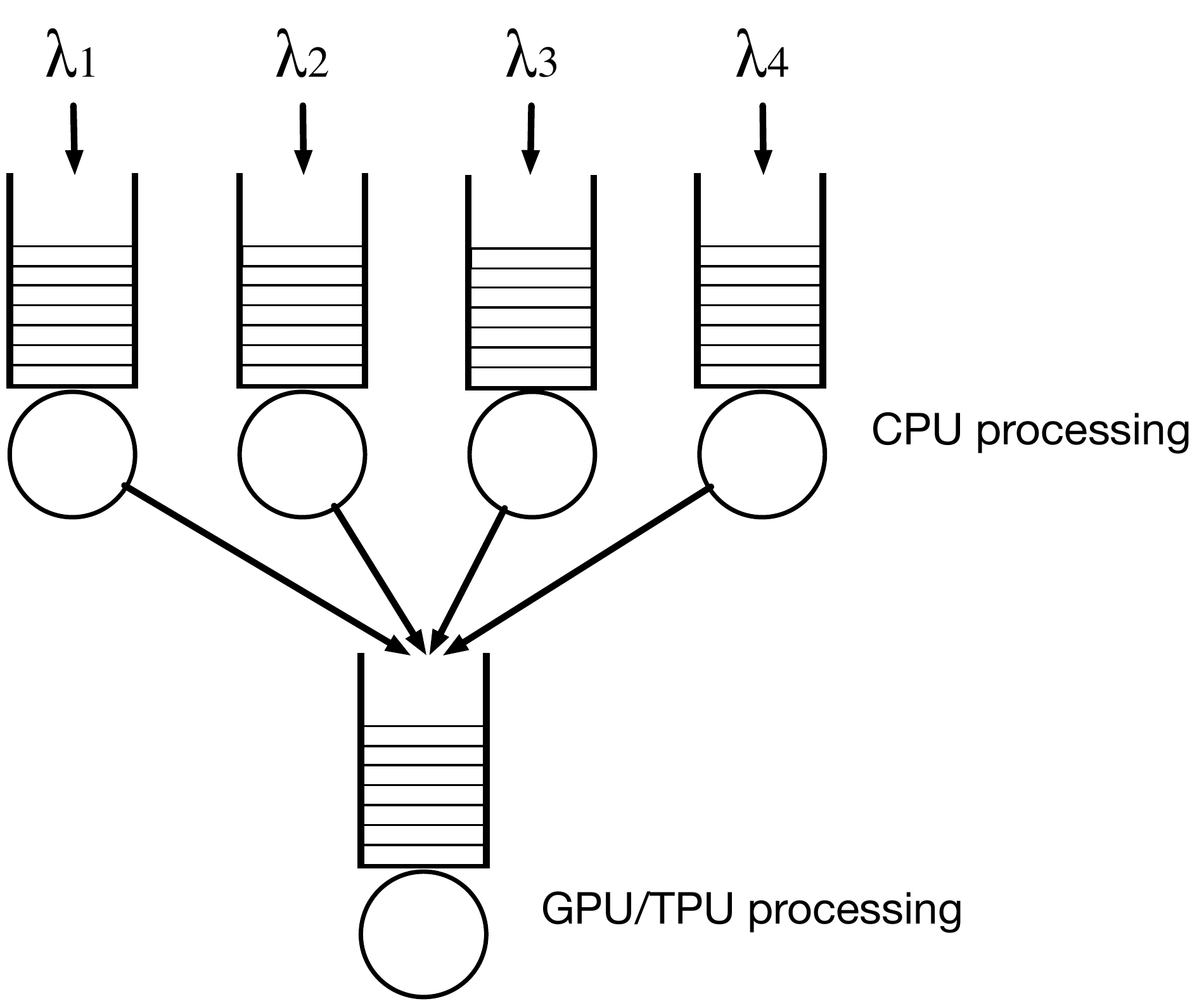}\label{fig:tandem_queue}
	}
		\captionsetup{belowskip=0pt, aboveskip=0pt}

	\caption{%(a) Two tier architecture for DNN inference. 
	(a) Tandem queue model for  CPU   and  GPU/TPU processing of a request. (b)   Network of queues model showing one CPU queue per application and a single GPU/TPU queue for all applications.}
	%(a)Architecture view of multiple scenarios (b) Network of queues model for our two tier architecture. (c) Each frontend container is modeled as a separate queue to capture CPU processing. A single backend queue models the aggregate behavior of all backend containers resident on a node.
	\label{fig:network_queue}
\end{figure*}

We model this CPU and GPU/TPU processing using a network of queues model, with separate queues to capture the CPU and GPU/TPU processing of a request (see fig \ref{fig:network_queue}). In the simplest case where a single application runs on the edge server (i.e., $k=1$), this model reduces to a tandem queue shown in fig \ref{fig:tandem_queue_arch}. In the general case where $k$ applications are co-located on an edge server, we model each application's CPU processing as a separate queue. In contrast, the GPU/TPU processing of all  applications is modeled as a single queue that is fed by the $k$ CPU queues. Fig \ref{fig:tandem_queue} shows the network of queueing model. We make this design choice since CPU processing of the $k$ co-located applications is \emph{isolated} from each other (since applications run inside containers or virtual machines and CPU is a virtualized resource with isolation). Edge accelerators, in contrast, lack hardware support for virtualization. As a result, edge GPU and TPU processing is not isolated, and GPU/TPU requests from all $k$ applications will be multiplexed on to the accelerator without isolation. Hence, we model GPU/TPU processing as a single queue that services the aggregate GPU/TPU workload $\lambda = \sum_{i=1}^k \lambda_i$ of all $k$ applications. 

By using a shared queue, our analytic model can capture the performance interference between applications on the accelerator and its impact on per-application response time. Given this network of queues model, we next derive closed form equation of response times for TPU, GPU, and CPU processing for each queue for each queue in the network.

%\begin{figure*}
%\centering
%\begin{minipage}[t!]{0.7\linewidth}
% Two tiered Arch
%    \centering
	%\subfloat[]{
%		\includegraphics[width=0.5\textwidth]{imgs/inference_as_a_service.pdf}\label{fig:inference_as_a_service}
%	} \hspace*{0.2in}
%	\subfloat[]{
%		\includegraphics[width=0.35\textwidth]{imgs/user_trained.pdf}\label{fig:user_trained}
%	}
%	\caption{Two-tiered edge application architecture. (a) AI-as-a-service with multiple frontend containers shares a single backend. (b) User-trained models use a one-to-one mapping of frontend and backend containers.}
%	\label{fig:two_tier_arch}
%\end{minipage}
%\quad
%\begin{minipage}[t!]{0.25\linewidth}
%% Tandem Queue
%\centering
%\includegraphics[width=0.52\linewidth]{imgs/two_tier.pdf}
%\caption{Tandem queue represents a server running a single 2-tier application.}
%\label{fig:tandem_queue_arch}
%\end{minipage}
%\end{figure*}

\subsection{Modeling TPU Inference Processing}\label{sec:model_tpu}

We first model an edgeTPU accelerator that executes inference requests from $k$ co-located applications. We assume that all $k$ DNN models are loaded onto the TPU. When application $i$ receives an inference request, it invokes the $i^{th}$ DNN model for executing that request. Incoming TPU requests from all $k$ applications are queued up in a single shared queue and processed in FCFS order. TPU request processing is sequential  and non-preemptive in nature. Once TPU begins processing a request, the processing cannot be preempted. Upon completion, the DNN model corresponding to the next queued request is loaded from host RAM into device memory, resulting in context switching overhead before DNN inference can begin; no context switching overhead is involved if the next request invokes the same model as the previous one. 

\begin{figure*}
\centering
\captionsetup{justification=centering}
\begin{minipage}[t]{0.45\linewidth}
    \centering
    \includegraphics[width=\textwidth, valign=t]{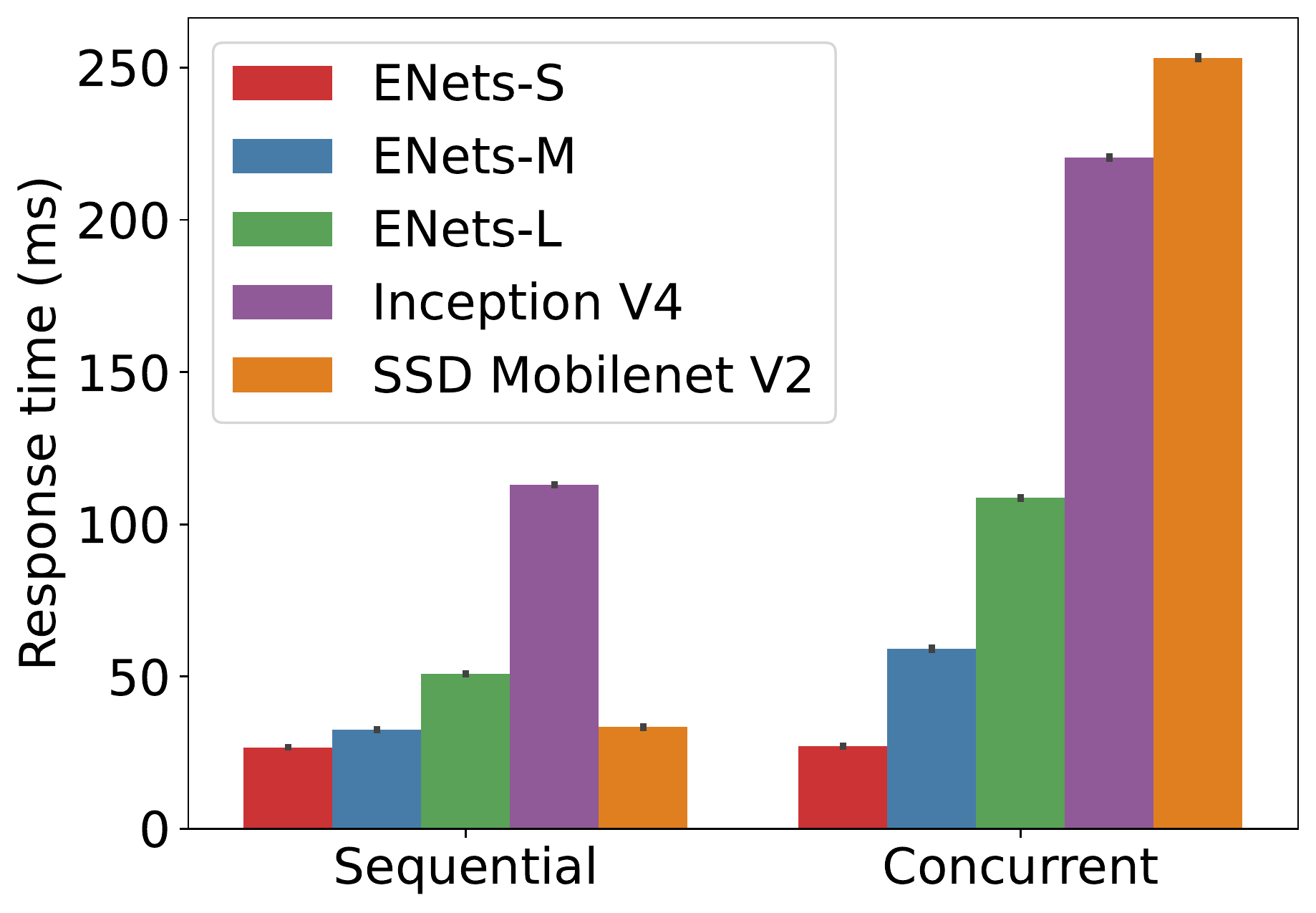}
    	\captionsetup{belowskip=0pt, aboveskip=10pt}

	\caption{Effect if Concurency on EdgeTPU}
	\label{fig:edgetpu-fcfs}
\end{minipage}
\hspace{0.4in}
\begin{minipage}[t]{0.45\linewidth}
    \centering
    \includegraphics[width=\textwidth, valign=t]{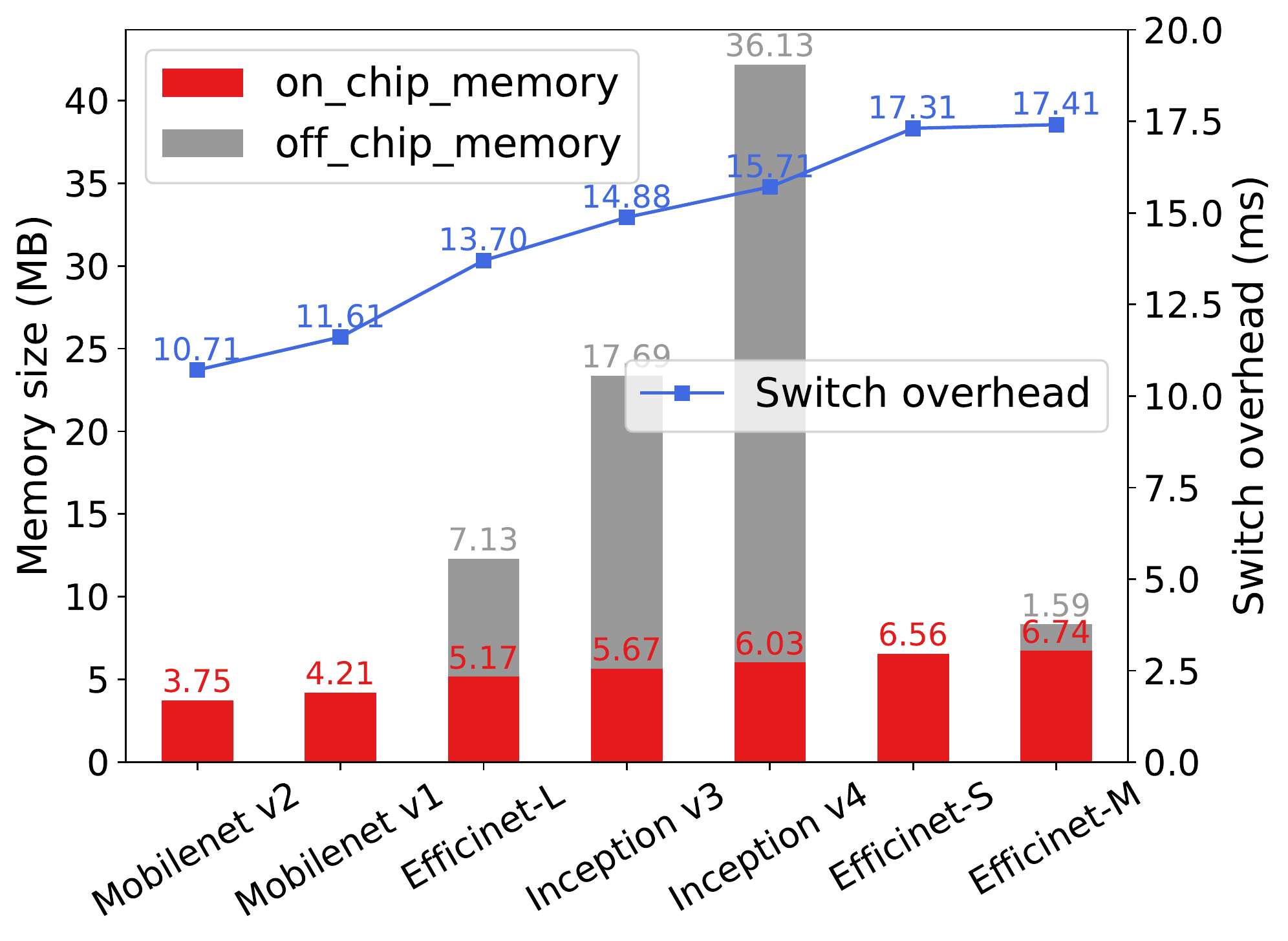}
    	\captionsetup{belowskip=0pt, aboveskip=0pt}

	\caption{EdgeTPU context switch overhead for various models.}
	\label{fig:tpu_switch_overhead}
\end{minipage}
\end{figure*}

To demonstrate that request multiplexing on an edgeTPU is FCFS and non-preemptive, we experimentally ran five different DNN models on an edgeTPU node. We first ran each DNN model by itself in isolation and then with all DNNs executing concurrently.  Fig \ref{fig:edgetpu-fcfs} depicts  the TPU execution (service) time in each case. As can be seen, in the presence of concurrent arrivals, the response time of requests beyond the first arrival includes the service time of previous arrivals, indicating FCFS and non-preemptive service.  In addition, there is a non-negligible context switching overhead when loading a new DNN model to the edgeTPU, making it important to model. We experimentally quantify this overhead in 
Figure \ref{fig:tpu_switch_overhead} in \S \ref{sec:tpu_validation}.
%later (in Section~\ref{sec:DNN_profiling}) show an experiment to quantify this overhead. 

%\begin{figure*}
%\centering
%\begin{minipage}[b]{0.57\textwidth}
%% Three models
%    \subfloat[EdgeTPU]{
%		\includegraphics[width=0.5\textwidth,valign=t]{imgs/edgetpu_fcfs}\label{fig:edgetpu-fcfs}
%	}
%	%\hfill
%	\subfloat[Jetson Nano]{
%		\includegraphics[width=0.5\textwidth,valign=t]{imgs/trt_ps}\label{fig:nano-ps}
%	}
%	\caption{Response time of sequential and concurrent requests.}
%\end{minipage}
%\hspace{0.1in}
%\begin{minipage}[b]{0.4 \textwidth}
%\includegraphics[width=\textwidth]{imgs/gpu_char.pdf}
%	\caption{GPU multiplexing behavior example}
%	\label{fig:gpu_example}
%\end{minipage}
%
%
%\end{figure*}
%

%\begin{figure}
%	\centering
%	\includegraphics[width=0.4\textwidth]{imgs/gpu_char.pdf}
%	\caption{GPU multiplexing behavior example}
%	\label{fig:gpu_example}
%\end{figure}

Therefore, when modeling edgeTPUs, the analytic model needs to consider  three important characteristics. 1) The device processes concurrent requests in an FCFS manner. 2) Context switching overhead to a different model is not negligible. 3) The mean service time across all requests seen by the TPU will be the weighted sum of the service times of requests of co-located DNN models. Hence, we model the queue for a TPU as an $M/G/1/FCFS$ queueing system, where arrivals are Poisson with a rate $\lambda$, the service times have a general distribution, and requests are scheduled in an FCFS manner.

 If $k$ applications run on the edge server, where $k \geq 1$, and each sees an arrival $\lambda_i$, then $\lambda = \sum_i \lambda_i$ denotes the aggregate request rate at the TPU. Let $S_i$ denote the expected service time, $e_i$ denote the execution time and $o_i$ be the switch overhead for workload $i$. Then we have
\begin{equation}
	S_i = P(M_i)e_i + (1-P(M_i))(e_i+o_i),
	\label{eq:tpu-service}
\end{equation}
where $P(M_i) = \lambda_i/\sum_k \lambda_k$ is the probability that a randomly chosen request in the system runs DNN model $i$.  Then the mean service time of the aggregated workload is the weighted sum of service time, weighted by arrival rates:
\begin{equation}
	S = \sum_i P(M_i)S_i
\end{equation}
%\sout{Note that we need to add context switch overhead to execution time when estimating $S_i$ (i.e. The modeled service time includes the context switch overhead).} 
For an $M/G/1/FCFS$ queueing system, the mean queueing delay seen by requests is given by the well-known Pollaczek-Khintchine (P-K) formula \cite{mor, gross2008}:
\begin{equation} \label{eq:p-k}
	E[w] = \frac{\rho+\lambda\mu \mathrm{Var}[S]}{2(\mu-\lambda)}
\end{equation}
where $\mu = 1/S$ is the TPU service rate, $\rho=\lambda/\mu$ is the utilization, and $\mathrm{Var}[S]$ denotes the variance of service time $S$. In the special case where there is a single tenant on the TPU, the DNN execution times can be modeled as a deterministic process, which reduces the $M/G/1$ model to a $M/D/1/FCFS$ model. In this case, the waiting time for a $M/D/1/FCFS$ system is well-known and is given by:
%it is also possible to use a simpler $M/D/1/FCFS$ queuing system to model the system. The waiting time for a $M/D/1/FCFS$ system is well-known and is given by:
\begin{equation} \label{eq:md1-fcfs}
	E[w] = \frac{\rho}{1-\rho} \cdot \frac{1}{2\mu}
\end{equation}
% \sout{In either case, the mean response time seen by the aggregate workload is given by }

%\begin{equation}
%	E[R] = E[w] + \frac{1}{\mu}
%\end{equation}
%\sout{and that seen by a particular backend container $i$ can be approximated as }

In either case, the mean response time of a particular model $i$ can be approximated as
\begin{equation}\label{eq:md1-fcfs-container}
	E[R_i] = E[w] + S_i
\end{equation}

\begin{figure*}
\centering
\captionsetup{justification=centering}
\begin{minipage}[t]{0.45\linewidth}
    \centering
	\includegraphics[width=\textwidth, valign=t]{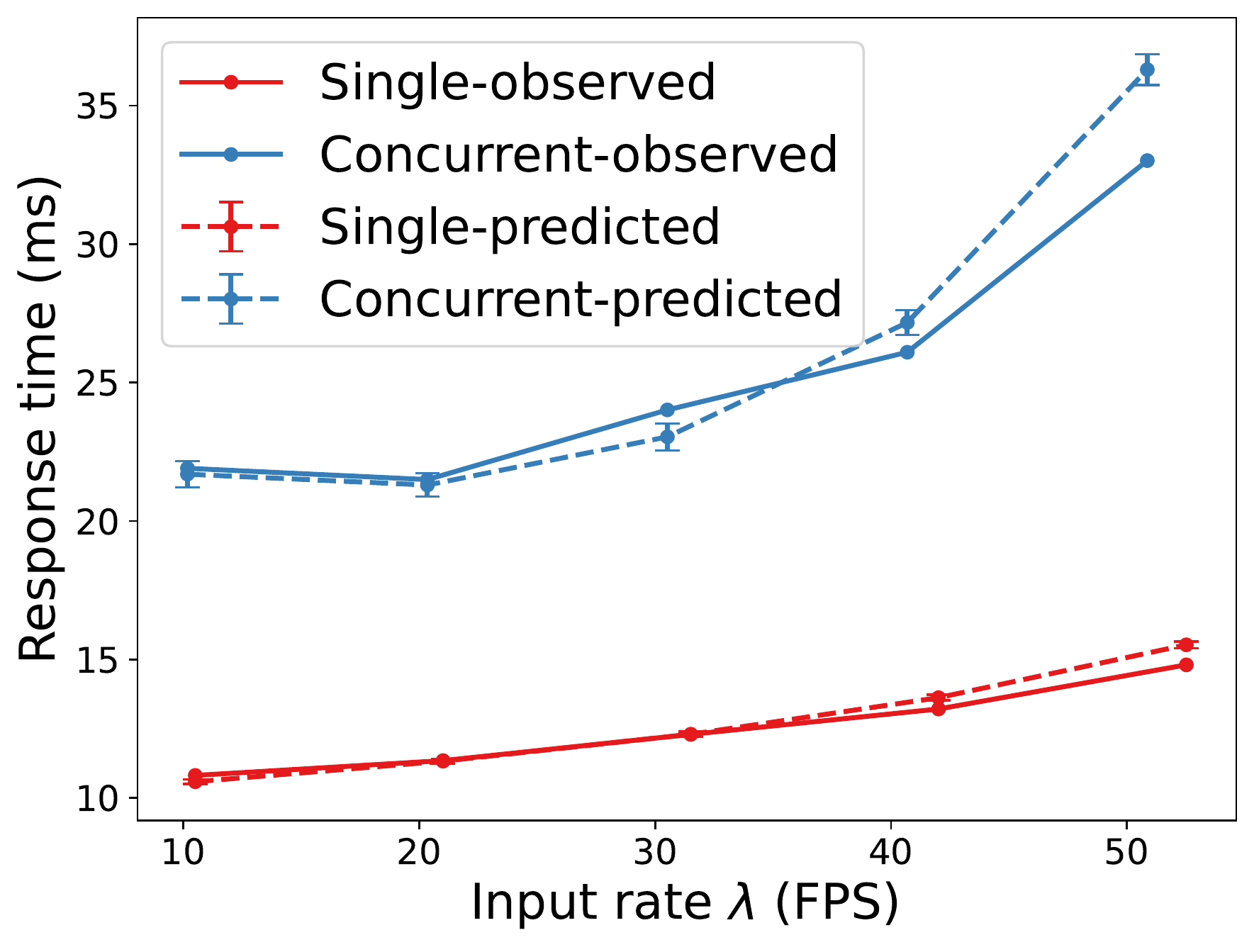}
		\captionsetup{belowskip=0pt, aboveskip=0pt}

	\caption{TPU response time of Efficientnet-S}
	\label{fig:edgetpu-queuing-single}
\end{minipage}%
\hspace{0.4in}
\begin{minipage}[t]{0.45\textwidth}
% TPU Validation
    \centering
	\includegraphics[width=\textwidth, valign=t]{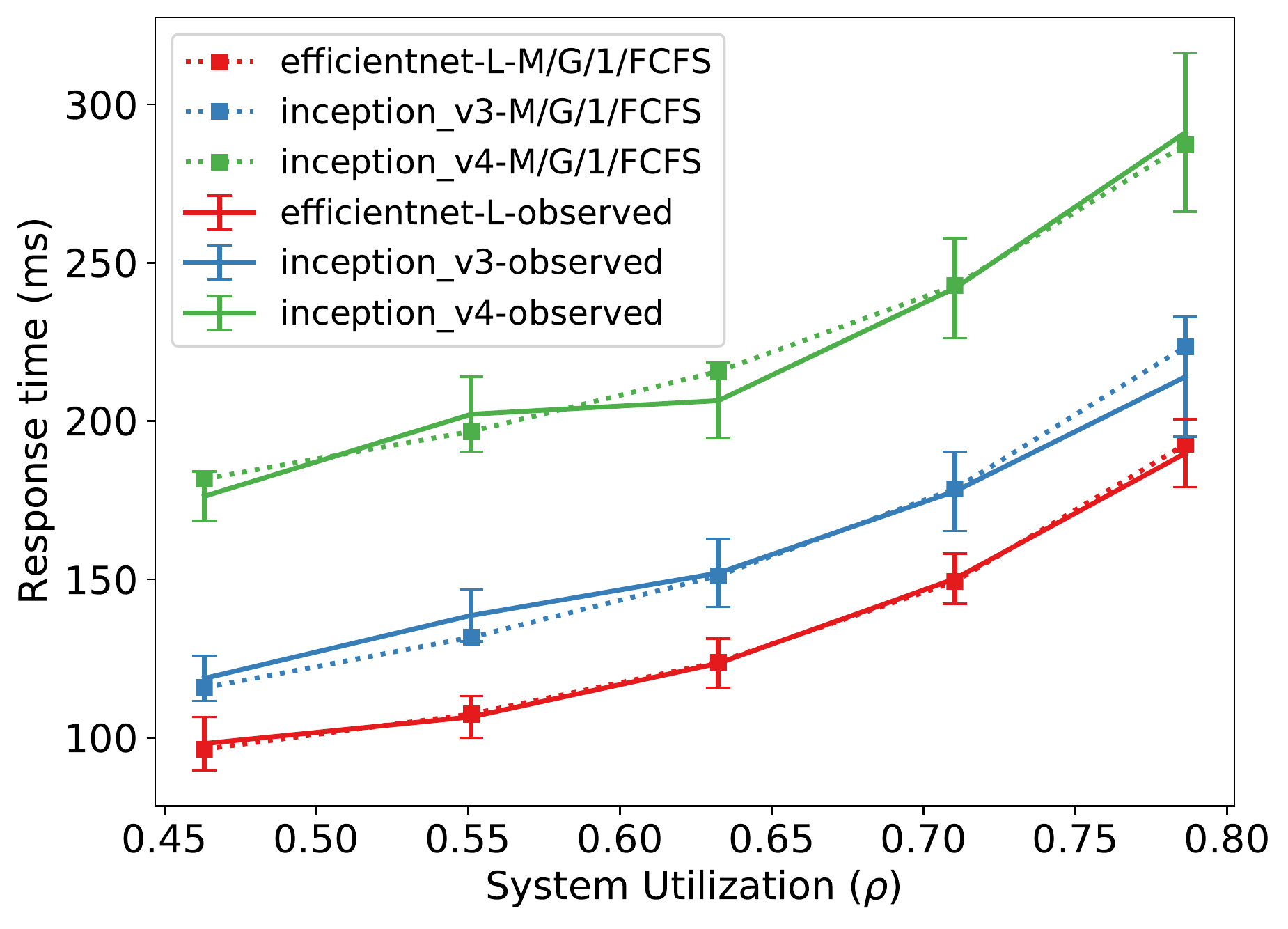}
		\captionsetup{belowskip=0pt, aboveskip=0pt}

	\caption{TPU response times for concurrent DNN models.}
	\label{fig:edgetpu-queuing-concurrent}
\end{minipage}%
\end{figure*}
\subsubsection{Experimental Validation of TPU Models} \label{sec:tpu_validation} 
We conduct an experimental validation of our analytic TPU model to demonstrate that it can accurately predict TPU response time for a broad range of real-world DNN models running concurrently on the edgeTPU.
To do so, we consider IoT applications that use different DNN models for image classification and object detection. We use two dozen DNN models from many of the most popular model families, namely, AlexNet, ResNet, EfficientNet, Yolo, Inception, SSG, VGG, and DenseNet. The characteristics of the used DNN models, along with their memory footprint and inference time are  summarized in Table \ref{tab:models} in Section \ref{sec:exp_resource_management}. The model sizes ranges from 3.5 million parameters to 144 million with a memory footprint of 22MB to 617MB. Collectively, these models range from small to large, both in their memory footprint and execution cost. 

%Our validation experiments in this and subsequent sections use 23 DNN models for image classification and object detection from many of the most popular DNN models. Table \ref{tab:all-models} in Appendix describes the characteristic of these 23 models.  

Our first experiment shows the context switch overhead when starting a new request execution. We loaded five DNN models on an edge server and sent a sequence of requests invoking these models in random order.  Figure \ref{fig:tpu_switch_overhead} shows the context switch overhead as well as 
on-chip and off-chip memory consumption of various models. Since the TPU stores all models in host server RAM and loads them to on-chip device memory on-demand, the context switch overhead is strongly correlated with model size (due to the overhead of copying the model from host RAM to TPU memory).  The context switch overhead ranges from 10 to 17ms, which is not negligible when compared to the DNN execution times that are shown in Figures \ref{fig:edgetpu-queuing-single} and \ref{fig:edgetpu-queuing-concurrent}. This experiment confirms the need to incorporate the context switch overhead when modeling request service times in Eq \ref{eq:tpu-service}.

%As shown, not all models fits on the TPU memory and the switch overhead is correlated with model size. These values has huge implications on device selection as we will see later. 

Next, we evaluate our TPU queueing model in the presence and absence of other co-located applications to demonstrate the impact of performance interference.  We run an application that uses the Efficientnet-S  model and vary its request rate.  Initially we run the application by itself on the edge server and measure the response time under different request rates. We then run the application along with a second application running Mobilenet-V1 that sees a constant workload. We measure the response times in the presence of this co-located application. Fig \ref{fig:edgetpu-queuing-single} depicts the observed response times and those predicted by our analytic models when running the Efficientnet-S by itself and with the background MobilenetV2 application. As can be seen, the response time seen by  Efficientnet-S is higher when it is running with a second application due to the performance interference from the background load. Further, our response times of our analytic model match closely with the observed response time, indicating that our model is able to capture the performance interference when sharing the TPU across multiple applications.

To further validate  our TPU model, we run various mixes of the DNN models on a TPU cluster node with varying request arrival rates. Fig \ref{fig:edgetpu-queuing-concurrent}  shows the response times seen by three co-located applications---Efficientnet-L, Inception-V3 and Inception-V4---under different levels of system utilization. In all cases, the analytic model predictions closely match the observed response times over a range of utilization values. Finally, we repeat the experiment with other mixes of DNN models and observed similar prediction accuracies (graphs omitted due to space constraints).

%Further, fig \ref{fig:edgetpu-queuing-single} shows the response time of Efficientnet-S when running by itself and fig \ref{fig:edgetpu-queuing-concurrent} shows the response time when it is colocated with the other two Mobilenet-V1 and Mobilenet-V2 models. The figure depicts how the response time of Efficientnet-S increases in the presence of other applications --- due to the lack of isolation and the resulting performance interference. In both cases, the model predicts closely match the observed response time over a range of utilization values. Finally, we repeat the experiment with serveral mixed of DNN model and observed similar behaviors (graphs omitted for space reason).

Overall, our validation experiments show that our analytic models can (1) accurately capture the context switch overheads and multiplexing behavior of the TPU, (2) can capture performance interference from co-located applications, and (3) can accurately predict TPU response time for real-world DNN models and workloads. 

% --------------------- OLD TEXT BEGIN ----------------------
%Figure \ref{fig:edgetpu-queuing} depicts one such combination where EfficientNet-L, Inception V3, and Inception V4 are co-located on a single TPU node. 
% The figure shows the observed TPU response times on the cluster closely matches those predicted by the queuing model  over a range of utilization values due to varying request arrival rates. We repeated this experiment with several different mixes of DNN models and observed similar behavior (graphs omitted for space reasons).. %Figure \ref{fig:edgetpu_queue_appendix} in appendix \ref{apx:model-valid} show a few selected combinations from our extensive set of experiments.
% --------------------- OLD TEXT END ----------------------

\subsection{Modeling GPU Inference Processing}\label{sec:model_gpu}

%Next, we model GPU behavior when processing concurrent requests. 

We next model an edge GPU accelerator that executes inference requests from $k$ co-located applications. 
Like in the TPU case, we assume that all $k$ DNN models are loaded onto the GPU, and applications issues requests to  these models upon receiving a request. Like before, all issued requests arrive at a shared queue and are processed by the edge GPU. Unlike the TPU that processes queued requests in an FCFS manner, edge GPUs, 
specifically those from Nvidia, have more sophisticated multiplexing capabilities. In particular, GPUs can process concurrent inference requests issued by different processes via {\em preemptive time-sharing} \cite{MPS}. That is, if $n$ concurrent requests are issued by $n$ independent applications, the GPU will serve the requests using round-robin time-sharing, where each request receives a time slice before being preempted \cite{MPS}. Unlike TPUs, which incur significant context switch overhead from memory coping, GPU's context switches are very efficient so long as models fit in GPU memory. Since GPUs have several GBs of on-device memory, we assume they can hold several models in RAM and are context switch overheads are negligible.

%Figure \ref{fig:gpu_example} illustrates the GPU's multiplexing behavior. The example assumes that two requests for DNN  A and one for DNN B arrive concurrently. The GPU processes one request each for DNN A and B using time-sharing. Upon the completion of DNN A's first request, it then processes the second request for DNN A along with the remainder of DNN B's request (the second request of DNN A needs to wait until the first one has finished since the CUDA context for DNN A processes multiple requests sequentially, based on CUDA behavior). 

\begin{figure*}
\centering
\captionsetup{justification=centering, position=b}

\begin{minipage}[t]{0.45\textwidth}
    \centering
    \includegraphics[width=\textwidth, valign=c]{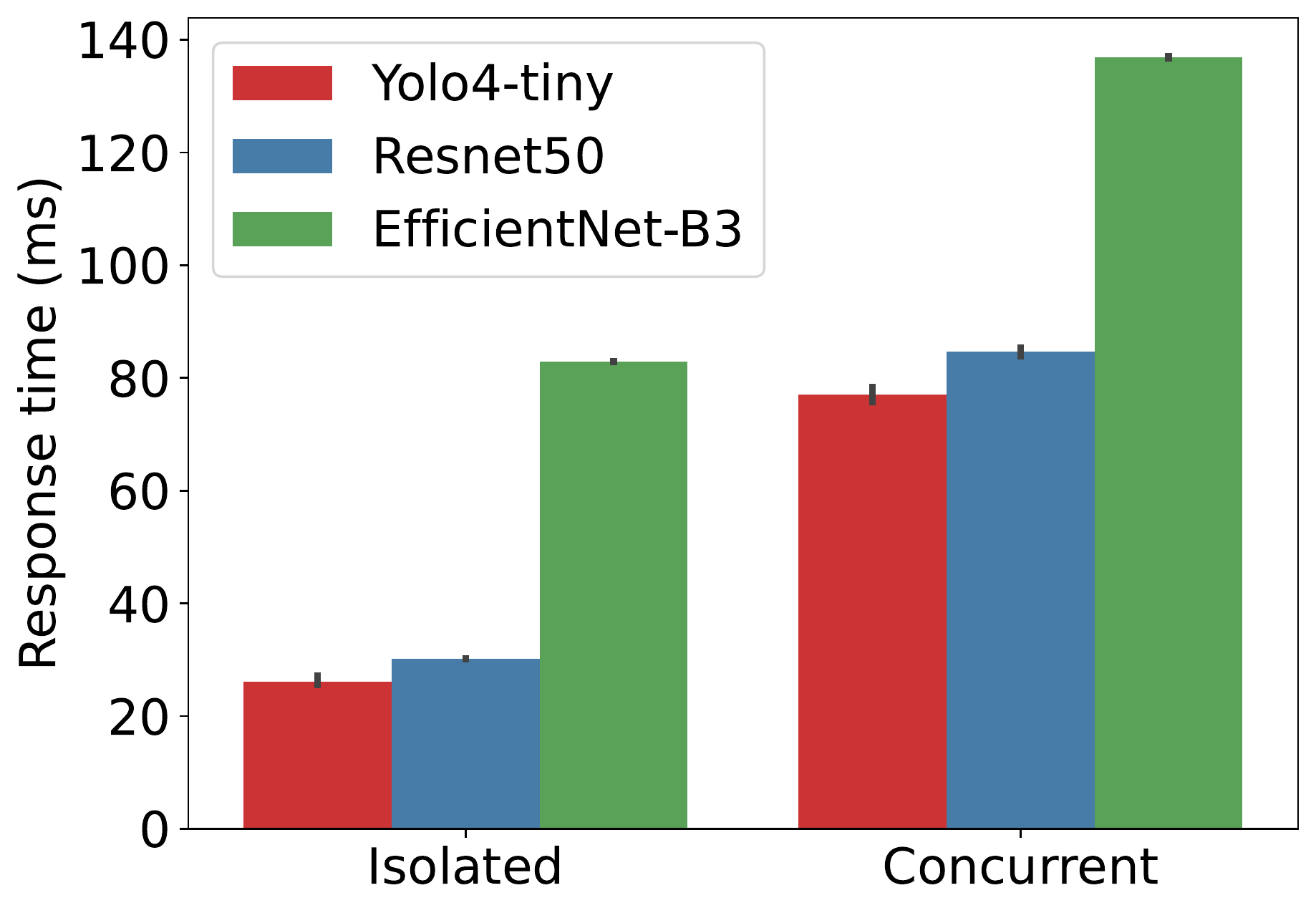}
    	\captionsetup{belowskip=0pt, aboveskip=5pt}

    \caption{Effect of Concurrency on Jetson Nano}
    \label{fig:nano-ps}
\end{minipage}%
\hspace{0.4in}
\begin{minipage}[t]{0.45\textwidth}
    \centering
	\includegraphics[width=\textwidth, valign=c]{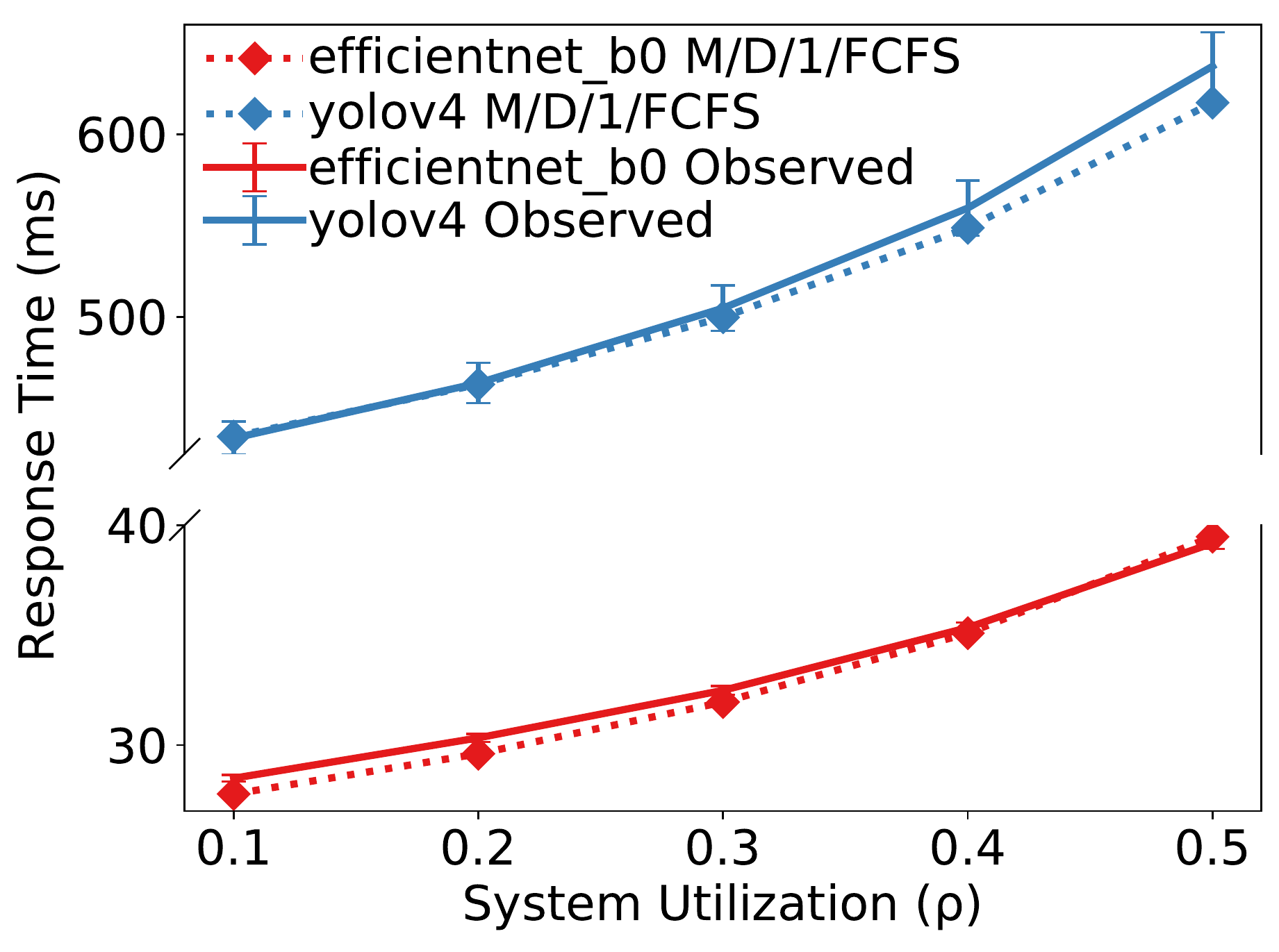}
	\captionsetup{belowskip=0pt, aboveskip=0pt}

	\caption{GPU response times for small and large DNN models.}
	\label{fig:nano_single_validation}
\end{minipage}
\end{figure*}

%We use same settings as section \ref{sec:model_tpu} except that we assume the models have already been loaded into GPU memory and thus no switch overhead. As shown in the figure, unlike EdgeTPU, GPU begins to process the model $B$ request as soon as it comes and alternate between model $A$ and $B$. As a result, the response time $R_A[0]$ of first request of model $A$ is 45ms and that of model $B$ is 30ms. Note that requests of the same model are queued by the application itself in FIFO manor. Thus, the second request of model $A$ will not be served until the first request is finished and the response time of the second request of model $A$ is 75ms.
To demonstrate time-sharing behavior of an edge GPU, we ran several DNN models on a Jetson Nano GPU, first in isolation and then  with all of them executing concurrently.   Fig \ref{fig:nano-ps} shows that when requests to various DNN models arrive concurrently, their completion times reflect time-sharing behavior. For example, requests to Yolo4 and Resnet DNNs, shown in red and blue, complete at nearly the same time, which can not happen if they were processed sequentially in an FCFS manner. This experiment confirms the time-sharing behavior of the GPU.

From a queueing perspective, the time-sharing capability of the GPU lends itself to a process-sharing (PS) queueing discipline. However, there are some important system issues to consider. First, despite the GPU's time-sharing capabilities, multiple requests issued by the \textit{same} process (i.e., application) are serviced in FIFO order. This is because the GPU associates a single CUDA context to each OS process, and all requests from a CUDA context go into a FIFO queue on the device and are serviced sequentially (FCFS fashion). Time-sharing is possible only when concurrent requests are issued by \emph{separate} processes (i.e., separate applications)) from distinct CUDA contexts. 

Since we  model the workload from all applications using a single queue, the resulting behavior will resemble some combination of an FCFS and PS queueing system. If multiple requests arrive concurrently from the same edge application, 
%different frontend containers arrive at the \emph{same } backend container, 
they will see FCFS processing. Conversely, if different applications issue concurrent requests to the GPU, these requests will see concurrent time-sharing processing (i.e., PS behavior). 

Put another way, if an incoming request sees an idle system, it experiences FCFS processing. If the GPU is busy processing a request and another request arrives at the \emph{same} application, it will be queued, also yielding FCFS behavior. In contrast, if the GPU is busy and a new request arrives at a \emph{different} application, all requests receive service via process-sharing (i.e., time-sharing).

We model this GPU behavior using a combination of $M/G/1/FCFS$ and $M/G/1/PS$ system, which serve as the upper and lower bounds of what requests actually experience in the system. The waiting time and response time for $M/G/1/FCFS$ are given by the P-K formula and are the same as  Equations \ref{eq:p-k} and \ref{eq:md1-fcfs-container}. The response time of a $M/G/1/PS$ queueing system has the following closed form solution \cite{mor}
\begin{equation}\label{eq:mg1-ps}
	E[R] = \frac{1}{\mu-\lambda}
\end{equation}
where $\lambda=\sum_i\lambda_i$ and $S=\frac{\sum_i\lambda_i S_i}{\sum_i \lambda_i}$ and $\mu=1/S$. Note that $M/G/1/PS$ has a well-known insensitively property, where the behavior is independent of job size distribution, which yield the same response time solution of $M/M/1/FCFS$ \cite{mor}. The mean queueing delay is 
\begin{equation}
	E[w] = E[R] - E[S] = \frac{1}{\mu-\lambda} - \frac{1}{\mu} 
\end{equation}
The application-specific response time can be approximated as:
\begin{equation}\label{eq:mg1-ps-container}
	E[R_i] = E[w] + S_i
\end{equation}

%\begin{figure}
%	\centering
%	\includegraphics[width=0.6\textwidth]{imgs/1080_cpu_validation.pdf}
%	\caption{CPU Model Validation}
%	\label{fig:cpu_validation}
%\end{figure}
%\begin{figure}
	%\centering
	%\includegraphics[width=0.6\textwidth]{imgs/edgetpu_queuing_validation}
	%\caption{Example of observed response time vs. predited response time when loadding three models, Efficientnet-L, Inception V3 and Inception V4, on one EdgeTPU device.}
	%\label{fig:edgetpu-queuing}
%\end{figure}

\subsubsection{Experimental Validation of GPU Model}

We now experimentally validate the above GPU model to show that it can accurately predict GPU response time and capture the multiplexing behavior of the GPU under different conditions. We also show that the real-world edge GPU behavior is bounded between our FCFS and PS models depending on arrival patterns.  To do so, we first run a single DNN model on the GPU and subject it to various arrival rates. We choose  an isolated EfficientNet\_b0 (a small classification DNN)  for this experiment and then repeat it with an isolated YoloV4 (a large object detection DNN) in isolation. In either case, since only a single CUDA context is running in the application, the request processing will be FCFS.  Figure \ref{fig:nano_single_validation} shows the GPU response times at different utilization levels compared to the queueing models predictions.  As shown, the  observed GPU response times closely match the values predicted by our FCFS  GPU model. This validates the FCFS behavior of the device under concurrent requests from the same application and also the ability of our model to capture this FCFS behavior.

\begin{figure*}%
    \centering
    \subfloat[EfficientNet B1 x ResNet 50 ]{{
        \includegraphics[width=0.32\linewidth]{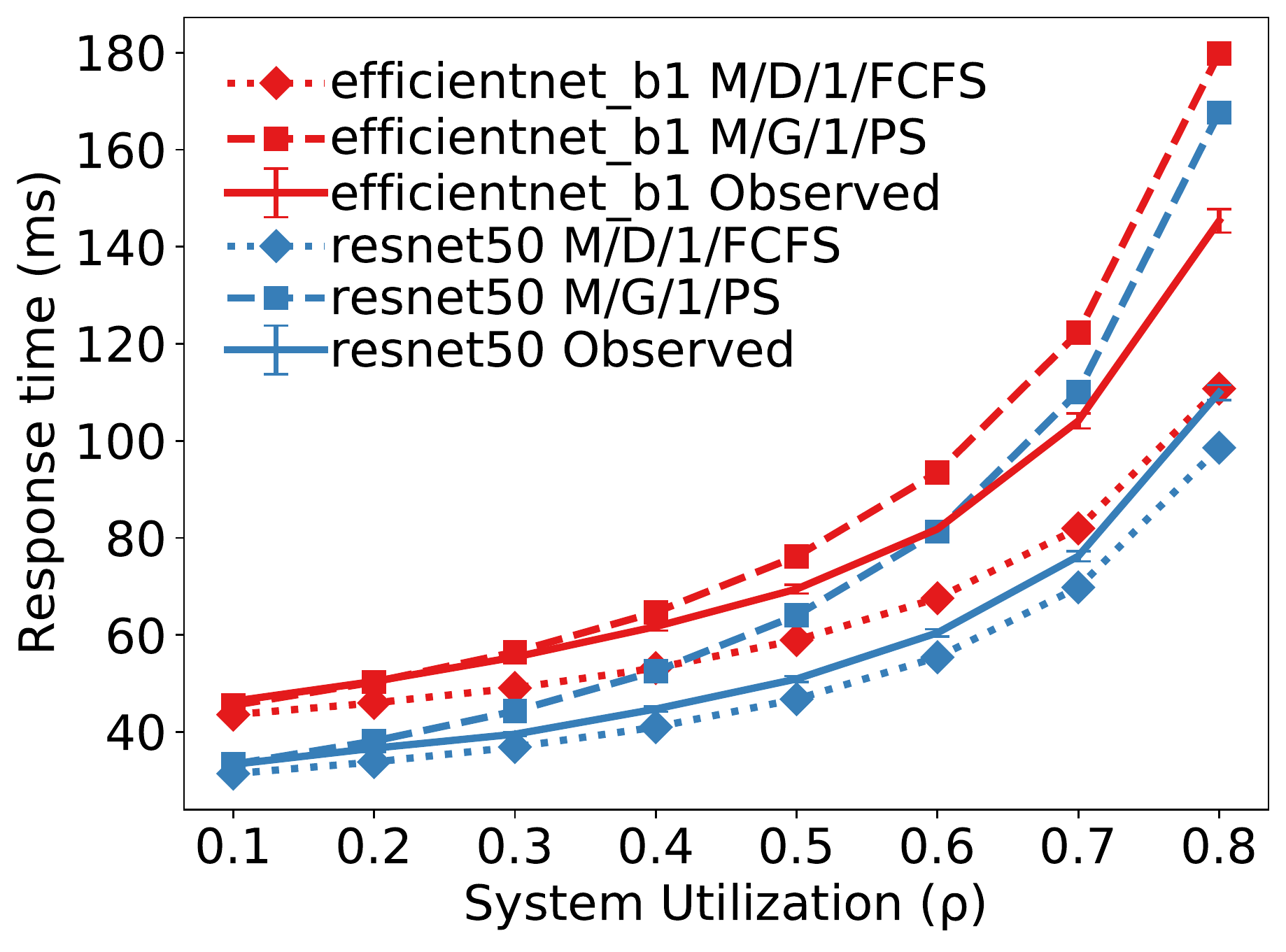} 
        \label{fig:nano_mp_small}
    }}%
    \subfloat[Three Models]{{
        \includegraphics[width=0.32\linewidth]{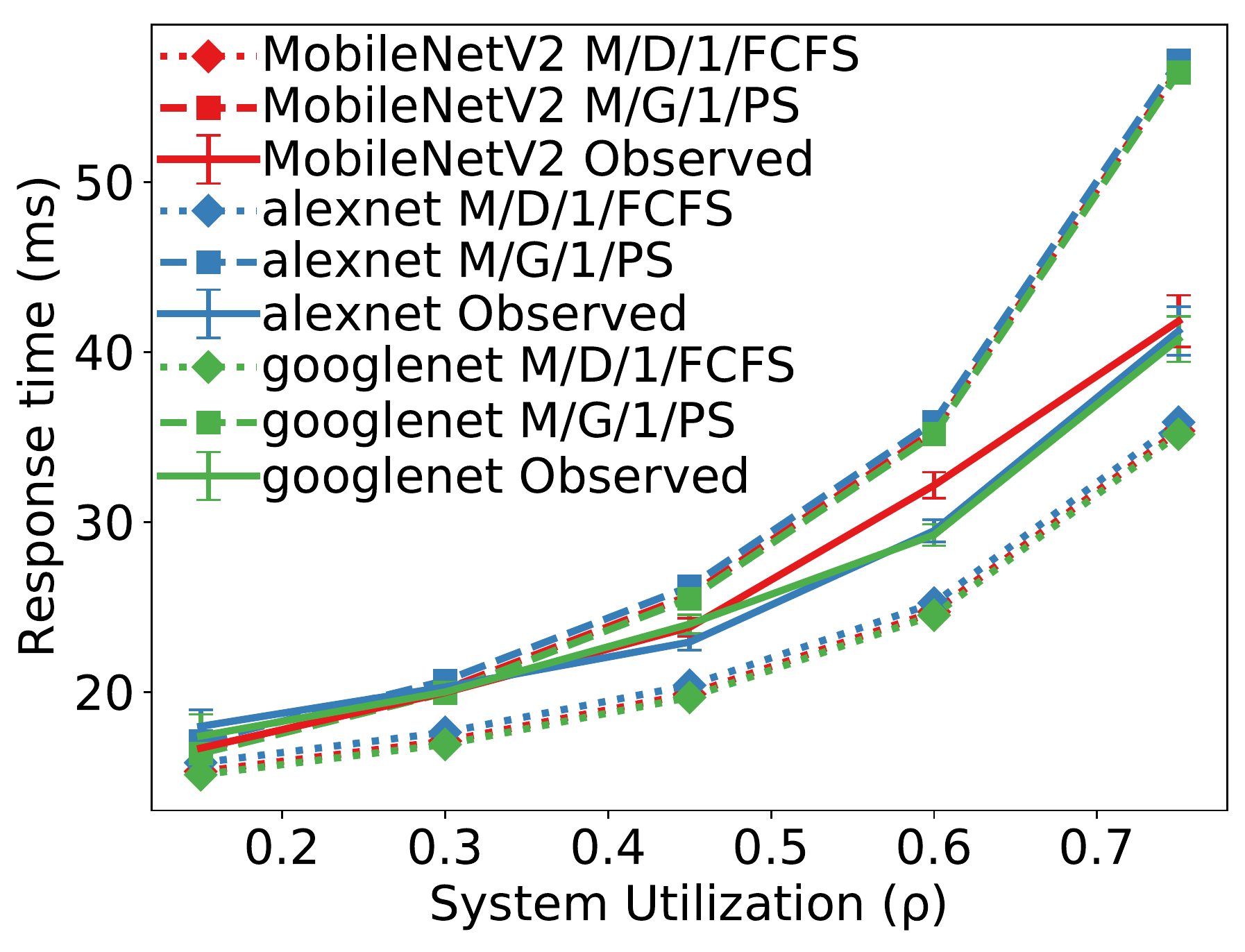} 
        \label{fig:nano_three_models}
    }}%
    \subfloat[Enforced PS and FCFS Behaviour]{{
    \includegraphics[width=0.32\textwidth]{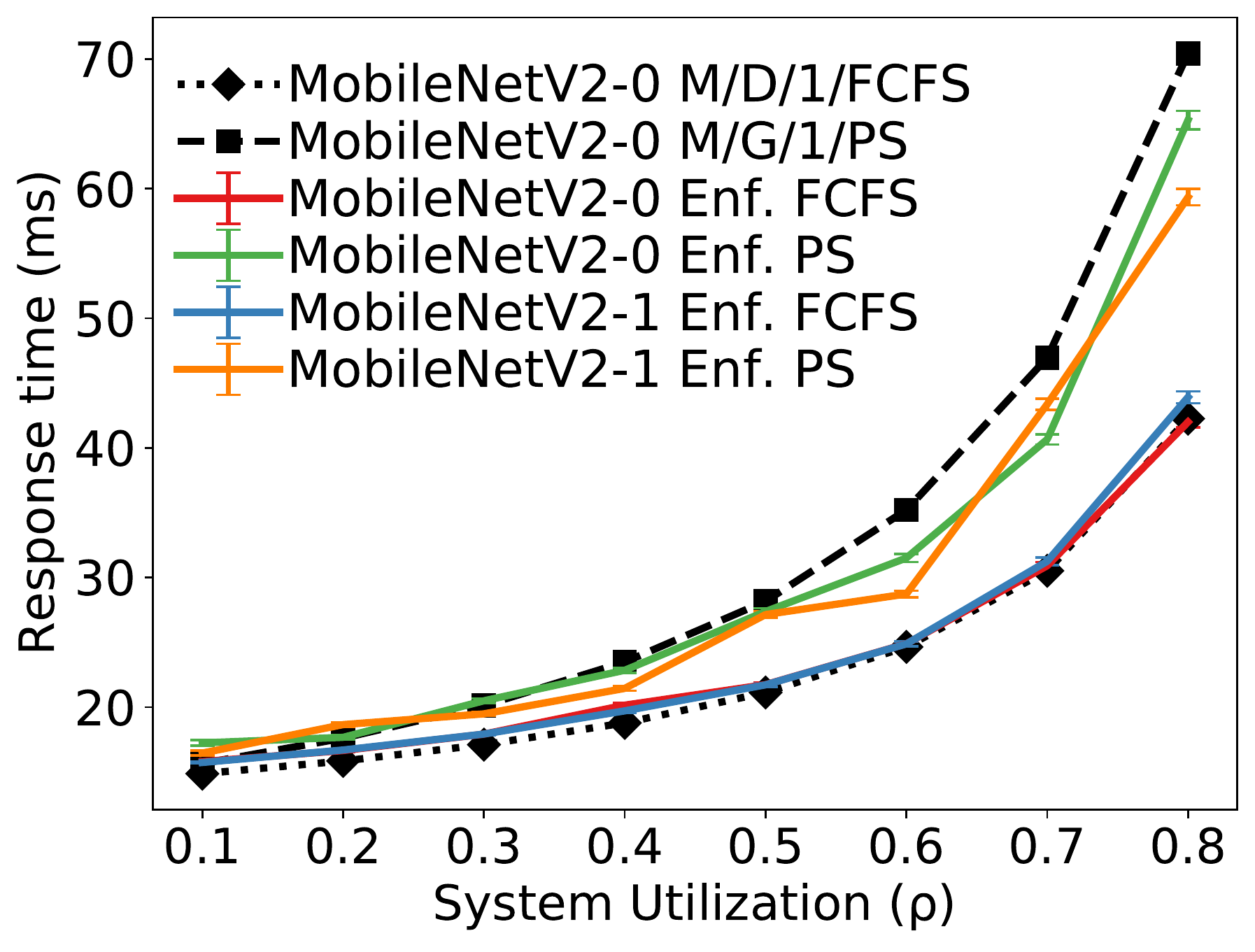}
	\label{fig:nano_two_enforced}
    }}%
    \caption{GPU behaviour with multiple applications per node.}%
    \captionsetup{belowskip=-30pt, aboveskip=-10pt}
\label{fig:nano_multiprocess_validation}%
\end{figure*}

Next, we validate our queueing models when multiple models run on a single node. Here, the observed response time should be lower and upper bounded by the pure FCFS and PS queueing models.  
Figure \ref{fig:nano_mp_small} shows the predicted response times according to the queueing models along with the observed response times when running two concurrent models on the cluster, namely, ResNet 50 and EfficientNet\_b1. In addition, 
Figure \ref{fig:nano_three_models} depicts a node with three concurrent DNN models (MobileNetV2, AlexNet and GoogleNet). In all cases, we see that the observed response time curve for each model lies {\em between} the PS and FCFS model curves. When a request arrives to an idle system or to the same application that served the previous request, it experiences FCFS behavior. Concurrent requests to different DNN models see time-sharing behavior.

To further demonstrate the accuracy of the queueing models, we run an experiment where we use a node with two MobileNetV2 models and force FCFS scheduling by generating arrivals with no execution overlap. We then repeat the run forcing PS behavior using perfectly synchronized arrivals, hence causing time-shared processing.  The results of the experiment are shown in Figure \ref{fig:nano_two_enforced}. For the former, the observed response curve matches the FCFS model curve; In the latter case, the curve shifts down and closely matches the PS model curve. This  validates our assumption that the Nano GPU exhibits a mix of both behaviors.

Together, these experiments show that our models can capture the capture the time-sharing behavior of the GPU and  also predict GPU response times in the presence of concurrent applications.% of GPU inference processes between the upper PS and lower FCFS bounds. 

% Two Models

\subsection{Modeling Parallel Inference Processing} 
\label{sec:model_concurrency}

While time-sharing is the default multiplexing behavior of a GPU, GPUs also support two types of parallelism, which we model in this section. First, GPUs support batch processing, where a batch of requests is issued to the GPU and data parallelism is used to process the batch in parallel. Second, discrete GPUs support request parallelism, where GPU cores are used to process multiple requests in parallel. We present models to capture both data parallelism and request parallelism multiplexing behaviors.

\subsubsection{Modeling Batch Inference} \label{sec:model_concurrency_batch}
Consider a GPU device that receives a batch of $b$ requests for inference. Batch inference processing exploits data parallelism to distribute the processing of multiple requests in a batch across device cores. Consequently, processing the $b$ requests on a single batch is faster than processing them individually (as $b$ separate requests). From a modeling perspective, batch inference can be viewed as an increase in the GPU processing capability due to the speedup seen by batch inference---based on data parallelism.
Consequently, we can model batch inference processing by viewing the batch of $b$ requests as a single logical request that sees a faster service rate $\mu_b$ than the service rate $\mu$ seen by individual requests. Similar to \cite{Nexus}, which reported this behavior as well, we model this faster service rate by estimating the service time of the logical request comprising the batch as follows:

%% Model By walid

\begin{equation} \label{eq:Sb_estimate}
    S_b = k_1 + \frac{k_2}{b} \  \mbox{where} \ b \geq 1
\end{equation}
where, $k_1, k_2$ are DNN model and device dependent constants, $b$ denotes the batch sizes. The constants $k_1, k_2$ are estimated empirically for each model. This model captures typical device behavior where the latency decreases initially with increasing batch size, followed by asymptotically diminishing improvements in the latency. Consequently, we model service time of a batch $S_b \propto 1/b$. The service rate $\mu_b$ is then $1/S_b$. The adjusted service time and service rate can then be used in the $M/G/1/FCFS$  system  from  \S \ref{sec:model_tpu}.

\subsubsection{Experimental Validation of Batched Inference}

We conduct  experiments to validate data parallelism with batched requests. We consider three models GoogleNet, InceptionV3, and DenseNet. We execute each DNN model individually on the Geforce-GTX-1080 GPU and empirically profile each model. To do so, we run each  DNN with  batch sizes of $b=1,2,4,8$, and find the parameters for $k_1$ and $k_2$ of equation \ref{eq:Sb_estimate}. Next we run each model on the GPU and vary the batch size from $b=1$ to $b=32$. Figure \ref{fig:batch_size_estimate} compares the model predicted service time from equation \ref{eq:Sb_estimate}, and the observed service time. Again, the model predictions closely match the observed response times  over a range of batch sizes. The figure also shows that service  time reductions show 
diminishing returns with the increase in batch size and the most significant gains are seen at small batch sizes. This is in line with our analytic model, which assumes asymptotically diminishing latency improvements with increasing batch size.

\subsubsection{Modeling Parallel Inference using MPS} \label{sec:model_concurrency_MPS}
In contrast to batching, which provided data parallelism, Nvidia's multi-process service (MPS) provides request parallelism when executing concurrent requests from different DNN models in parallel on different GPU cores. MPS also supports memory partitioning and isolation of GPU resources \cite{MPS}

The behavior of MPS can be modeled as a $M/G/c/PS$ system, with the GPU providing $c$ servers that 
can execute $c$ DNN models in parallel.  Assuming an aggregate arrival rate $\lambda$ across all GPU 
containers, a mean service time $S$ and service rate $\mu$, the response time yielded by a $M/G/c/PS$ system is given as 
\begin{equation}\label{eq:mps-model}
E[R] = \frac{c}{\lambda}.\frac{\rho}{1-\rho} = \frac{c}{c\mu - \lambda}
\end{equation}
In practice, the degree of multiprocessing depends on the (i) the actual number of cores on the device, 
and (ii) the model size in terms of its processing and memory needs. In most cases, with MPS enabled, the GPU acts as a multi-processor with a small value of $c$ (e.g., $c$ is often 2 or 3 for desktop-class GPUs, indicating a limited degree of parallelism).

\subsubsection{Experimental Validation of MPS-based Parallel Inference} 

\begin{figure*}
\centering
\captionsetup{justification=centering}
\begin{minipage}[t]{0.32\linewidth}
% Batch Estimate
    \centering
    \includegraphics[width=\textwidth]{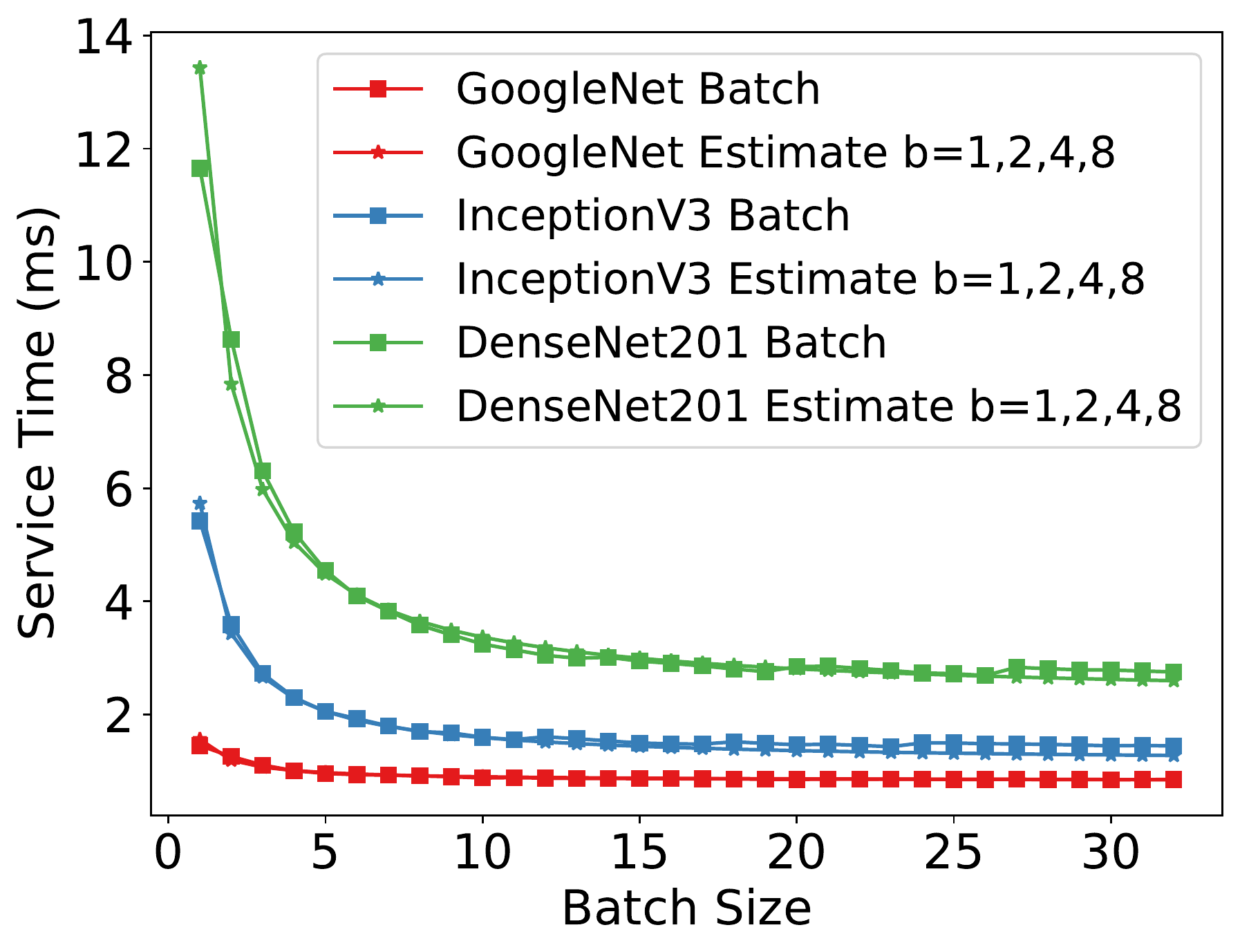}
    	\captionsetup{belowskip=0pt, aboveskip=5pt}

	\caption{GPU response times for batched request execution.}
	
	\label{fig:batch_size_estimate}
\end{minipage}%
\begin{minipage}[t]{0.32\linewidth}
% MPS
    \centering
    \includegraphics[width=\textwidth]{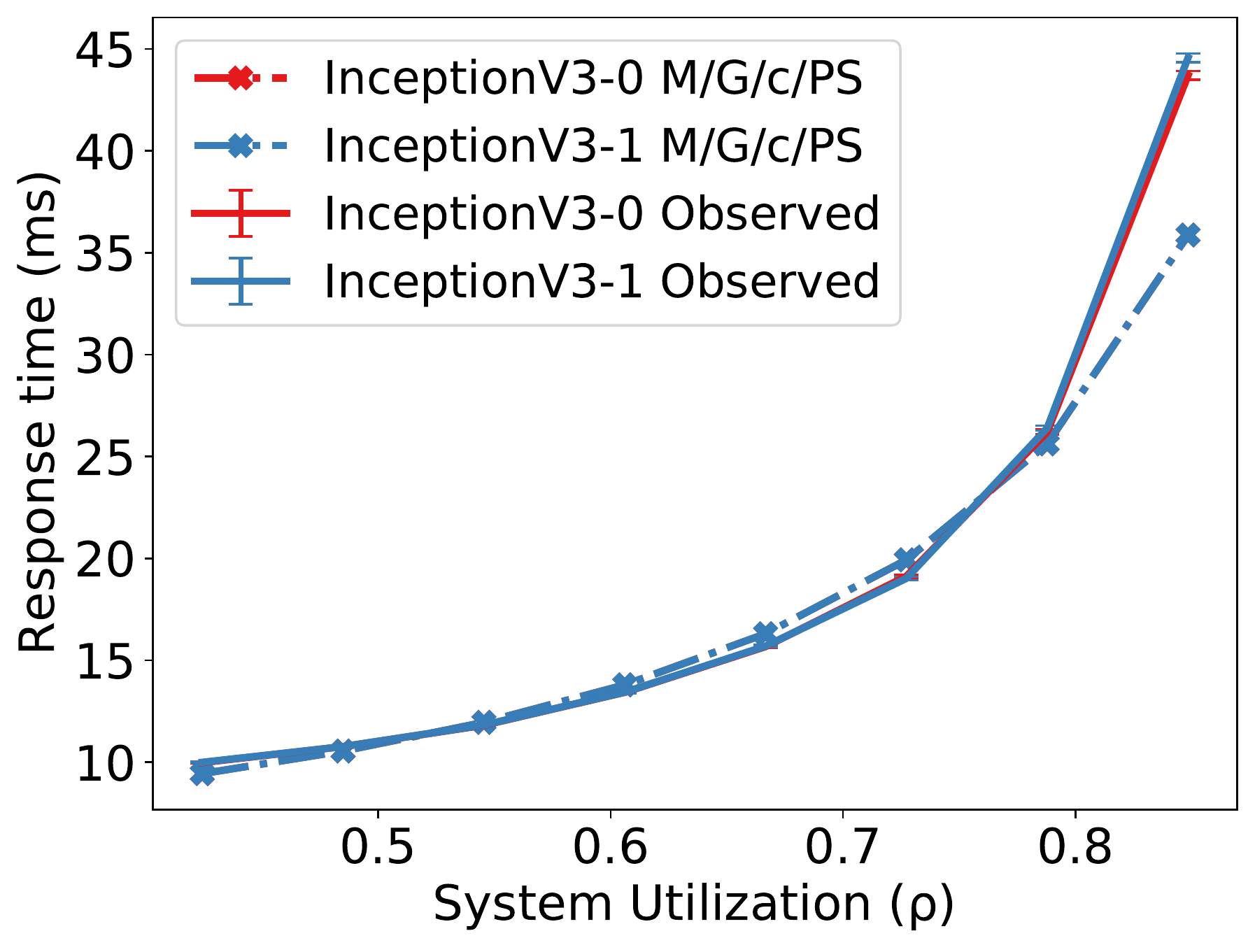} 
    	\captionsetup{belowskip=0pt, aboveskip=5pt}

    \caption{Response time under MPS for two DNNs with c=1.65.}%
    \label{fig:mps_multiprocess_validation}%
\end{minipage}%
\begin{minipage}[t]{0.32\linewidth}
% CPU Validation
    \centering
	\includegraphics[width=\textwidth]{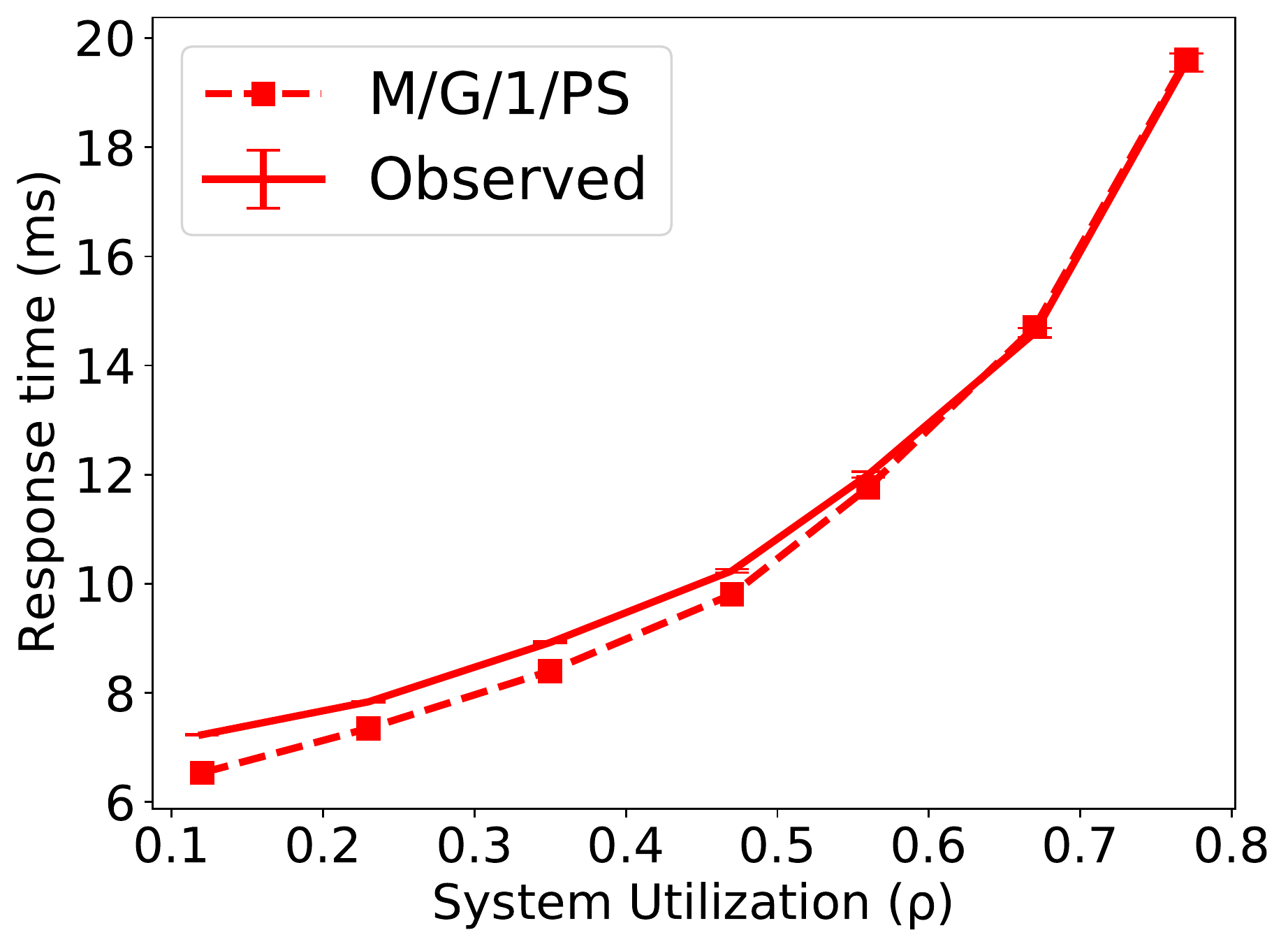}
		\captionsetup{belowskip=0pt, aboveskip=5pt}

	\caption{CPU response times at different utilization levels.}
	\label{fig:cpu_validation}
\end{minipage}
\end{figure*}

We conduct experiments to validate the above model that captures request parallelism due to Nvidia's MPS scheduler. In this case, we ran two InceptionV3 models on the GeForce 1080 GPU with the MPS daemon running with each having a varying request arrival rate. Figure \ref{fig:mps_multiprocess_validation} compares the observed response time to those predicted by the queueing model  for $c=1.65$ (empirically measured speed up). As shown, our model has a good match with the observed values, which shows that our analytic model is able to accurately capture parallel request processing on GPUs.

\subsection{Modeling CPU processing} \label{sec:model_cpu}

%An IoT application's frontend  container  1) listens for inference requests on a network socket, 2) performs any preprocessing for each incoming request, and 3) issues the request to the backend tier for inference. 

Having analytically modeled GPU and TPU processing under a range of multiplexing behaviors such as FCFS, time-sharing, and data/request parallelism. We turn to CPU processing incurred by each edge request. Unlike GPU and TPU accelerators, where we use a single queue to model the aggregate workload of all $k$ applications, the CPU model assumes a separate queue for each application. This is because each application is assumed to run in a separate container or a VM on the edge server and is  allocated a dedicated amount of CPU capacity (e.g.,  a fraction of a CPU core or multiple cores). We assume that the operating system uses standard CPU time-sharing for  processing requests within each container or VM.  

Under this scenario, the CPU processing of each application can be modeled as a separate $M/G/c/PS$ queueing system. 
%Within queuing theory, a notation such as $M/G/c/PS$ is used to describe (i) the probability distributions of request arrivals, (ii) the distribution of request service time, (iii) the number of processors and (iv) the scheduling discipline use to execute requests. Here, we assume Markovian arrivals (denoted by $M$), which means arrivals follow a Poisson distribution with a rate $\lambda$.  Request service times are assumed to have any arbitrary (i.e., general) distribution (denoted by $G$). The container is allocated $c$ cores, $c\geq 1$.  Requests are assumed to be scheduled using processor sharing (``PS'') to capture time sharing CPU scheduling.  We assume that the container has a maximum processing capacity (i.e., service rate) of $\mu$ requests/s based on the allocated CPU capacity. In any queuing system with arrival rate $\lambda$ and service rate $\mu$, the system utilization $\rho$ is given as $\rho = \lambda/c\mu$.  
The $M/G/c/PS$ queueing system is well studied in the literature and has a closed form equation for average response time, which is given by \cite{mor}:
\begin{equation} \label{eq:cpu-model}
E[R] = \frac{c}{\lambda} \cdot \frac{\rho}{1-\rho} = \frac{c}{c\mu - \lambda} \mbox{, since }\rho = \lambda/c\mu
\end{equation}

\subsubsection{CPU Model Validation}
%In Section \ref{sec:model_cpu}, we modeled the CPU behavior of the frontend containers as an $M/G/c/PS$ queuing system. 

To experimentally validate our CPU model, we allocate a fixed amount of CPU and memory to each  application container and ran a multi-threaded process to accept incoming image requests and perform CPU processing on this request prior to issuing it to the GPU or TPU
% steps, such as image resizing, to match the input size expected by the backend DNN model. 
 We varied the arrival rate and experimentally measured the  response time for various degrees of CPU utilization. Figure \ref{fig:cpu_validation} shows that the observed CPU response time   increases with utilization and  closely matches the prediction of our model, showing that the model accurately captures the CPU processing of the application.  Results for multi-core containers $(c>1)$ are similar and omitted due to space constraints. Our results show that our model can  capture CPU processing of DNN inference requests and that use of separate queueing models for each applications yields accurate response time predictions.

%\[E[R] = \frac{c}{\lambda} \cdot \frac{\rho}{1-\rho}\]
% The previous section used queuing models to capture the behavior of inference processing on a GPU or a TPU. In this section, we analyze the CPU processing an inference request performed by the frontend container.

%Let us assume there are $u$ frontend containers correspond to $u$ users. The $u$ containers are mapped to $n$ backend containers. The containers see an input arrival rate of $\lambda_i^f$, and each request incurs a mean service time $S_i^f$.

%Let $c$ denote the number of processing CPU core are on the edge server we model a system as a \textbf{M/G/c/PS} system where arrival rate are poisson. The aggregate arrival rate is $\lambda=\sum \lambda_i^f$, service time $S = \sum \lambda_i^f S_i^f / \sum\lambda_i^f$ and $\mu=1/S$. The total response time is 
%\begin{equation}
%E(r) = \frac{c}{\lambda}.\frac{\rho}{1-\rho}
%\end{equation}
%The CPU and inferences process can be viewed as a tandem queue where the request first undergoes CPU processing followed by inference processing in the GPU/TPU.
%
%The total response time is the sum of the two $E(r) = E(r_f)+E(r_b)$
%

\subsection{Estimating End-to-End Response Time}
Our previous sections presented analytic models for the CPU and GPU/TPU processing stages for each inference request. To estimate the end-to-end response time, we can use the network of queues model from figure \ref{fig:network_queue}. The mean end-to-end is the sum of the mean response time of each stage in the network. For application $i$ the end-to-end response time $R_i^{total} = R_i^{CPU} + R_i^{GPU/TPU}$, where  $R_i^{CPU}$ and $R_i^{GPU/TPU}$ can be computed using the above analytic models.

\section{Model-Driven Cluster Resource Management}
In this section, we show how the predictive capabilities of our analytic models can be employed for cluster resource management tasks such as online DNN placement and dynamic migration. We also discuss the implementation of our algorithms into our Ibis prototype that is based on the kubernetes cluster manager.
\RestyleAlgo{ruled}
\SetKwComment{Comment}{/* }{ */}

\begin{algorithm}[t]
	\caption{Latency-aware Online Knapsack Placement}
	\KwIn{A dictionary \texttt{dnnConfig} contains DNN profile, task type, input rate $\lambda$ and latency constrain $\tau$; A list of available nodes, \texttt{nodeList}.}
	\KwOut{A \texttt{selectedNode} to place the incoming DNN such that all applications running on the node will not violate their latency SLOs after placement}
	$\texttt{feasibleNodes} \gets []$ \;
	$\texttt{memNeed} \gets \texttt{computeMemNeed(dnnConfig)}$ \;
	\For{\texttt{node} in \texttt{nodeList}}{
		$\texttt{resTime} \gets \texttt{computeResponseTime(dnnConfig, node)}$ \Comment*[r]{Use queueing models}
		$\texttt{sysUtil} \gets \texttt{computeSysUtilization(dnnConfig, node)}$\;
		$\texttt{memOk} \gets \texttt{memNeed} < \texttt{node.freeMem}$\Comment*[r]{Check knapsack resources}
		$\texttt{sloOk} \gets \texttt{resTime} < \tau $\; 
		$\texttt{utilOk} \gets \texttt{sysUtil} < \texttt{maxRho}$\;
		\If{$\texttt{memOk}\ \&\&\ \texttt{sloOk} \ \&\& \ \texttt{utilOk}$}{
			$\texttt{feasibleNodes.append(node)}$\;
		}
		}
		\If{\texttt{dnnConfig.type} == "AIaaS"}{
			$\texttt{groupNodes} \gets \texttt{findNodeWithSameApp(dnnConfig, feasibleNodes)}$ \;
			$\texttt{feasibleNodes} \gets \texttt{groupNodes} \textbf{ if } \texttt{!groupNodes.empty()}$
		}
		$\texttt{selectedNode} \gets \texttt{findNodeWithLeastUtil(feasibleNodes)}$
		\Return \texttt{selectedNode}
	\label{alg:latency_aware}
\end{algorithm}
\subsection{Latency Aware Online Knapsack Placement}
A key resource management task performed by cluster managers in cloud computing platforms is online placement.\footnote{Offline placement  assumes that all applications arrive at once and must be placed together onto an empty cluster, while online placement assumes applications arrive sequentially and must be incrementally placed without knowledge of future arrivals.} In this case, new applications arrive into the cluster and must be placed onto a server with adequate unused capacity to run that application. The placement problem in cloud computing has been well studied for over a decade \cite{Urgaonkar2007ApplicationPO} and is typically viewed as a multi-dimensional knapsack problem \cite{knapsack}. In this case, each server in the cluster is a knapsack and the various dimensions of the knapsack represent resource capacities of the server (e.g., capacities of the CPU, memory, network). An application request specifies the amount of each resource needed by it and the knapsack problem involves selecting an edge server with sufficient unused resources.

In the case of edge clusters with accelerators, the traditional knapsack placement approach is not applicable. This is because traditional knapsack placement in cloud computing has assumed that resource requirements are additive and an application can run on a server if it "fits" on that server. This assumption is reasonable since all applications runs in  VMs or containers and are isolated from each other using virtualization.

As explained in \S \ref{sec:queue-models}, accelerators do not provide support for virtualization or isolation and co-located DNN applications on a GPU or TPU see performance interference. Hence, it is no longer sufficient for placement techniques to check if the DNN model will "fit" on a GPU or TPU memory. The placement technique should additionally ensure that the increased workload and performance interference from placing a new application will not cause  response time requirements to be violated. We refer to this new problem as \textit{online placement with latency constraint}, where the knapsack (i.e., edge server) has both resource capacity constraints as well as a response time (latency) property. While used resource capacity increases in an additive manner with each placed application, response time increases non-linearly. Thus traditional packing algorithms that are based on linear packing assumptions do not hold in our setting.

Our placement approach is based on our analytic queueing models to address the latency constraint. We assume that a newly arriving DNN application specifies  its resource (e.g., CPU, memory) needs as well as a latency constraint ($R_i^{th}$), for an expected request rate $\lambda_i$.

Our placement technique first determines a list of all feasible servers in the cluster that can house the new application. A server is feasible if (1) it has sufficient free resources such as CPU, CPU Memory, GPU memory to house the DNN application and (2) the end-to-end response time seen by the new as well as existing applications  are below their specified thresholds. That is, for each application on that server $R_i^{cpu} + R_i^{GPU/TPU} \leq R_i^{threshold}$, $R_i^{cpu}$ and $R_i^{GPU/TPU}$ are computed using our analytic models.

If no feasible server exits, the application placement request is rejected. Otherwise, a greedy heuristic is used to pick a specific server from the list of feasible candidate servers. Currently, our system supports two greedy heuristics. (1) highest utilization (aka worst fit) that is designed to achieve a tight packing on the smallest number of servers and (2) lowest utilization, which chooses the least loaded server to house the new application. Algorithm 1 lists the pseudo code for our online knapsack placement with latency constraints.

\subsubsection{Heterogeneous and Grouped Placement}\label{sec:placement_enhacement}
We next present two enhancements to our baseline placement algorithm, namely heterogeneous and grouped placement. The heterogeneous placement algorithm assumes an edge cluster with heterogeneous servers, where each server is equipped with a GPU, TPU, or both. The goal of the placement algorithm is to choose  a suitable server and the best accelerator type for a newly arriving application. To do so, the placement algorithm computes the CPU response time of the application on each feasible server as well as the GPU and TPU response times, depending on the accelerators on each feasible server. The placement algorithm greedily chooses the GPU server as well as the TPU server with the least latency from this feasible set.

The server with the lower of the two latencies is then chosen to house the application, which also determines the best accelerator  for the application. As we show in \S \ref{sec:DNN_profiling}, no one accelerator is optimal for all DNN models, and this enhancement enables the best accelerator to be chosen, based on the performance offered by each type of accelerator  and the load on servers.

Our second enhancement use a \emph{grouped} placement technique that we refer to as \emph{AI-as-a-Service} (AIaaS) placement. In the AIaaS model, the edge cloud provider offers a choice of several pre-trained DNN models as an edge service. An application can simply choose one of these DNN models and avoid having to supply its own DNN for inference tasks. The advantage of the AIaaS model is that the cluster manager can opportunistically group applications that choose the same DNN model on the same server --- in doing so, it can load a single copy of the DNN for all applications that have chosen it, instead of loading one DNN model per application. This can potentially increase the cluster capacity due to the reduced memory requirements. In the grouped placement, if the DNN model is already executing on the edge cluster (by being chosen by one or more prior applications), then we need to determine if the newly arriving application can be grouped with the existing ones. To do so, we aggregate the request rate $\lambda_i$ of all  grouped applications and use the analytic queueing model to determine the GPU and TPU response time of the shared container. If the latency threshold of the entire group can be met for the aggregate workload, the new application is co-located with the current group. If the application cannot be placed with the current group, it is placed onto a new server to start a new grouping.

\subsection{Model-driven Dynamic Migration}

Application workloads in edge clouds tend to be dynamic and will fluctuate over time. While our analytic model ensures that response times meet latency objectives for a specified request rate  $\lambda_i$, applications will nevertheless experience latency violations if the workload fluctuates dynamically and the observed   request rate $\hat{\lambda_i}$ exceeds the specified workload.  To handle dynamic workloads, Ibis uses a \emph{police-and-migrate} strategy. Each application container includes a token bucket regulator to police the incoming workload---the token bucket is configured with a rate $\lambda_i$ and configurable burst $b_i$. Hence, if the observed request rate exceeds the specified rate, requests get queued  by the token bucket regulator. Doing so avoids overloading the    underlying accelerator and isolates other co-located tenants from experiencing performance degradation due to the overloaded tenant. A sustained hotspot is mitigated by dynamically migrating the overloaded tenants to a new less-loaded server.  To do so, Ibis monitors the mean end-to-end  response time for each application over a moving time window. It also tracks the observed request rate $\hat{\lambda_i}$ for each application. If latency violations or request drops are observed on any node, the application whose observed request rate $\hat{\lambda_i}$ exceeds the specified rate $\lambda_i$ is flagged for migration.  The analytic queueing models are used to determine a new node and accelerator for this application using a new higher request rate estimate. The application is  then migrated to the new node using container or VM migration.

%Figure \ref{fig:cluster} shows the 10-node edge cluster used in our experiments. This cluster is consisted of 10 Nvidia Jetson Nano boards. We also use Jetson Nano as host of our EdgeTPU accelerators. As shown in the figure, 8 out of the 10 nodes are connected with an EdgeTPU through USB. 

\begin{figure*}
\centering
\captionsetup{justification=centering}
\begin{minipage}[t]{0.45\linewidth}
	\centering
	\includegraphics[width=\linewidth]{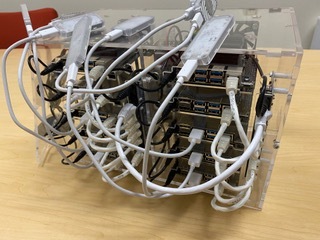}
		\captionsetup{belowskip=5pt, aboveskip=10pt}
	\caption{10-node Jetson Nano cluster}
	\label{fig:cluster}
\end{minipage}%
\hspace{0.4in}
\begin{minipage}[t]{0.45\linewidth}
    \centering
	%\captionsetup{belowskip=-14pt}
	\includegraphics[width=\linewidth]{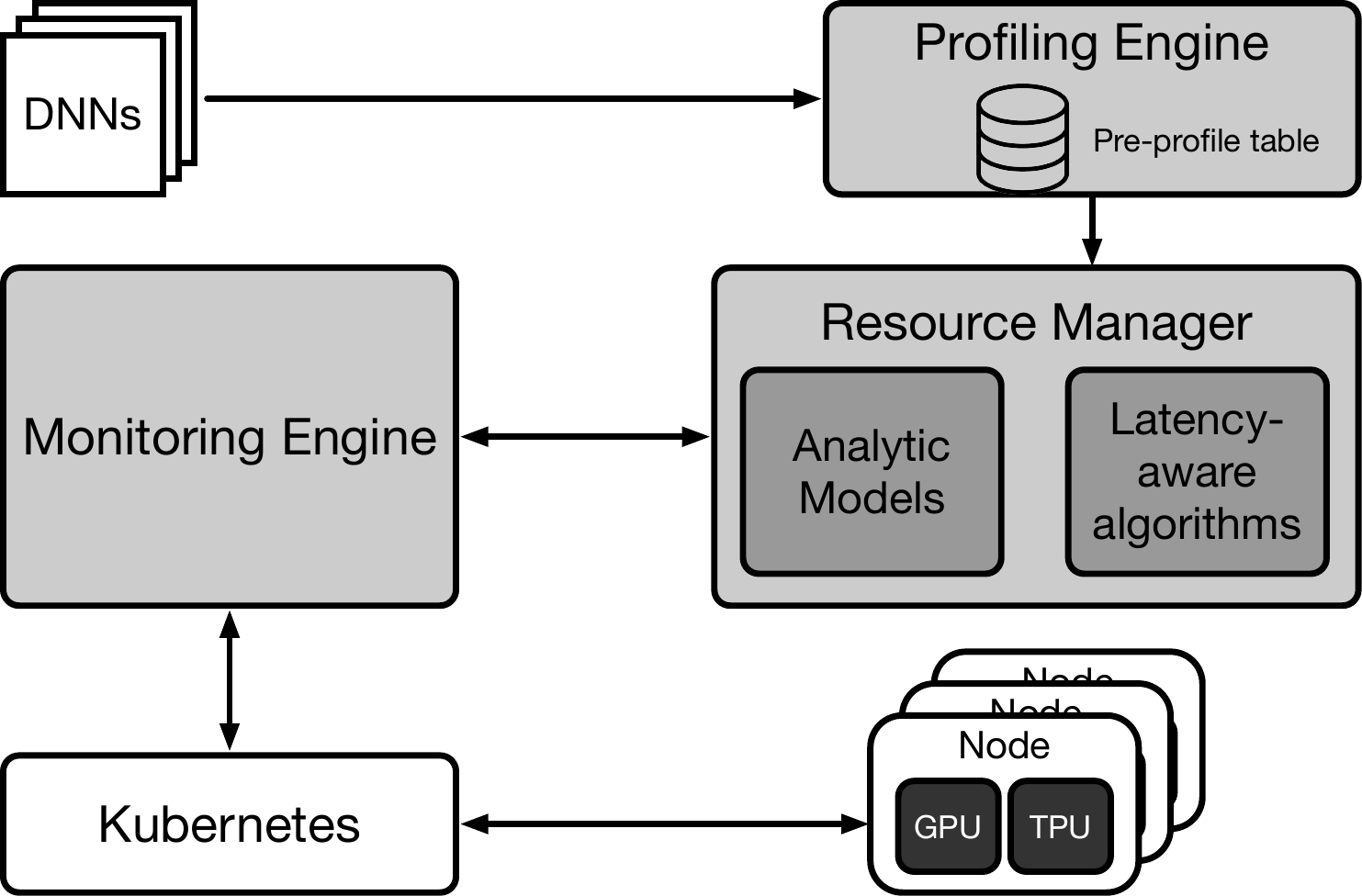}
		\captionsetup{belowskip=5pt, aboveskip=10pt}

	\caption{Ibis implementation overview}
	\label{fig:impl_overview}
\end{minipage}%
\end{figure*}  

\subsection{Ibis Implementation}

We have implemented a prototype of our system, named Ibis, using Kubernetes on a  custom-built edge cluster of ten nodes depicted in Figure \ref{fig:cluster}, each comprising a Jetson Nano GPU, 4 GB of RAM, and quad-core ARM processor. We use a similar ten node cluster equipped with USB edgeTPUs for our TPU experiment. Our cluster also contains a Dell PC with a 3.2 GHZ i7-8700 processor, 16 GB RAM, and Geforce-GTX-1080 GPU. All machines are connected via a gigabit ethernet switch and run Linux Ubuntu 18.04. Further, each Jetson Nano runs CUDA 10.2, cuDNN 8, and TensorRT 7.1.3 while the PC runs CUDA 11, cuDNN 7.6.5, and TensorRT 7.1.3. All machines are virtualized using Docker 19 and managed by Kubernetes 11; gRPC 1.31 is used for inter-communication.

Figure \ref{fig:impl_overview} depicts Ibis system overview. Our prototype currently runs DNN models in TensorRT engine format for GPU and quantized \texttt{tflite} format for EdgeTPU. For AIaaS models, Ibis first retrieves its resource requirements (e.g., memory, service time) using a pre-profile table. On the other hand, in the case of User-Supplier models (provided in ONNX format), we compile and profile it on an idle node and save the result back to the profile table for future use. Our application containers  are deployed as Kubernetes \texttt{Deployment}, and we use Kubernetes \texttt{Service} to route network traffic. Ibis collects node resource status and chooses a node and device using our analytic models and scheduling strategies. Finally, Ibis deploys incoming workload to target node and device via Kubernetes interface. Source code for our prototype is available on Github \emph{(URL blinded)}.

\subsection{Experimental Evaluation of Model-driven Resource Management Algorithms}\label{sec:exp_resource_management}
% Please add the following required packages to your document preamble:
% \usepackage{graphicx}
\begin{table*}[t!]
\centering
\resizebox{0.9\textwidth}{!}{%
\begin{tabular}{|c|c|c|c|c|c|c|c|}
\hline
Models &
  Scale &
  Input Shape &
  Parameters &
  \begin{tabular}[c]{@{}c@{}}Static Size \\ (MB)$^*$\end{tabular} &
  \begin{tabular}[c]{@{}c@{}}Runtime \\ Footprint  (MB)$^*$\end{tabular} &
  \begin{tabular}[c]{@{}c@{}}FLOPS \\ (GF)\end{tabular} &
  \begin{tabular}[c]{@{}c@{}}Inference \\ Time (ms)$^*$\end{tabular} \\ \hline
\multicolumn{8}{|c|}{Classification Models}                                         \\ \hline
AlexNet         & S & {[}N, 3, 224, 224{]} & 62M     & 138 & 992  & 0.7    & 14.18  \\ \hline
GoogleNet       & S & {[}N, 3, 224, 224{]} & 6 M     & 39  & 893  & 2      & 13.37  \\ \hline
InceptionV3     & M & {[}N, 3, 224, 224{]} & 24 M    & 81  & 836  & 6      & 30.56  \\ \hline
MobileNetV2     & S & {[}N, 3, 224, 224{]} & 3.5 M   & 22  & 1130 & 0.6    & 13.02  \\ \hline
ResNet18        & S & {[}N, 3, 224, 224{]} & 12 M    & 69  & 930  & 1.8    & 10.83  \\ \hline
ResNet34        & M & {[}N, 3, 224, 224{]} & 21.2 M  & 155 & 1044 & 3.6    & 19.51  \\ \hline
ResNet50        & M & {[}N, 3, 224, 224{]} & 26 M    & 106 & 965  & 3.8    & 29.2   \\ \hline
ResNet101       & L & {[}N, 3, 224, 224{]} & 44.5 M  & 247 & 1135 & 7.6    & 50.32  \\ \hline
EfficientNet-b0 & S & {[}N, 3, 224, 224{]} & 5.3 M   & 30  & 1168 & 0.4    & 26.03  \\ \hline
EfficientNet-b1 & S & {[}N, 3, 240, 240{]} & 7.8 M   & 42  & 1184 & 0.7    & 41.32  \\ \hline
EfficientNet-b2 & M & {[}N, 3, 260, 260{]} & 9.2 M   & 49  & 1196 & 1      & 49.58  \\ \hline
EfficientNet-b3 & M & {[}N, 3, 300, 300{]} & 12 M    & 77  & 1229 & 1.8    & 81.67  \\ \hline
EfficientNet-b4 & L & {[}N, 3, 380, 380{]} & 19 M    & 124 & 1042 & 4.2    & 166.15 \\ \hline
EfficientNet-b5 & L & {[}N, 3, 456, 456{]} & 30 M    & 180 & 180  & 9.9    & 337.44 \\ \hline
DenseNet121     & L & {[}N, 3, 224, 224{]} & 7.2 M   & 50  & 910  & 3      & 30.14  \\ \hline
DenseNet201     & L & {[}N, 3, 224, 224{]} & 20 M    & 103 & 964  & 4      & 89.11  \\ \hline
VGG16           & L & {[}N, 3, 224, 224{]} & 138 M   & 407 & 1275 & 16     & 86.36  \\ \hline
VGG19           & L & {[}N, 3, 224, 224{]} & 144 M   & 463 & 1333 & 20     & 99.19  \\ \hline
\multicolumn{8}{|c|}{Object Detection Models}                                       \\ \hline
YoloV3          & L & {[}N, 3, 416, 416{]} & 62 M    & 617 & 1501 & 65.88  & 190.24 \\ \hline
YOLO-tinyV4     & M & {[}N, 3, 416, 416{]} & 6.06 M  & 75  & 938  & 6.91   & 23.79  \\ \hline
YoloV4          & L & {[}N, 3, 608, 608{]} & 64.43 M & 445 & 1329 & 128.46 & 407.91 \\ \hline
\end{tabular}%
}
\caption{DNN characteristics for GPU (Values are based on Jetson Nano with FP16, different configuration might yield different values).}
\label{tab:models}
\end{table*}

In this section, we experimentally evaluate our cluster resource management techniques on an edge cluster using realistic workloads. To do so, we deploy Ibis on our 10 node edge cluster and use the  trace workloads discussed below for our experiments.

\subsubsection{Trace Workloads} The  trace workload for each IoT application is constructed using our two dozen DNN models (see Table \ref{tab:models}) and a sequence of images from the ImageNet dataset \cite{imagenet}.   Further,   we use the public Azure trace \cite{Azure-traces} to construct realistic mixes of containerized applications of varying sizes.  To do so, we analyze the distribution of VM sizes  (CPU utilization and memory footprint) 
in the Azure trace and use them to generate the utilization and model size of our DNN-based applications. We categorize VMs as small ($<$2GB), medium (2-8GB), and large ($>$8GB) and find that proportions of small, medium, and large applications are 47\%, 33\%, and 20\% respectively.   The mix of small, medium, and large DNN models housed on the edge cluster is chosen in the same proportion.  Finally, models in the same category are chosen evenly. 

We use this Azure trace to generate an arrival trace of DNN applications that require placement on our cluster. Each arriving application is chosen from the above DNN types and is associated with a response time constraint $R_i$, and worst-case request rate $\lambda$.

\subsubsection{DNN profiling} \label{sec:DNN_profiling} 
\begin{figure*}[t!]
\centering
\captionsetup{justification=centering}
\begin{minipage}{0.45\linewidth}
\centering
% Two models forced
\includegraphics[width=\textwidth]{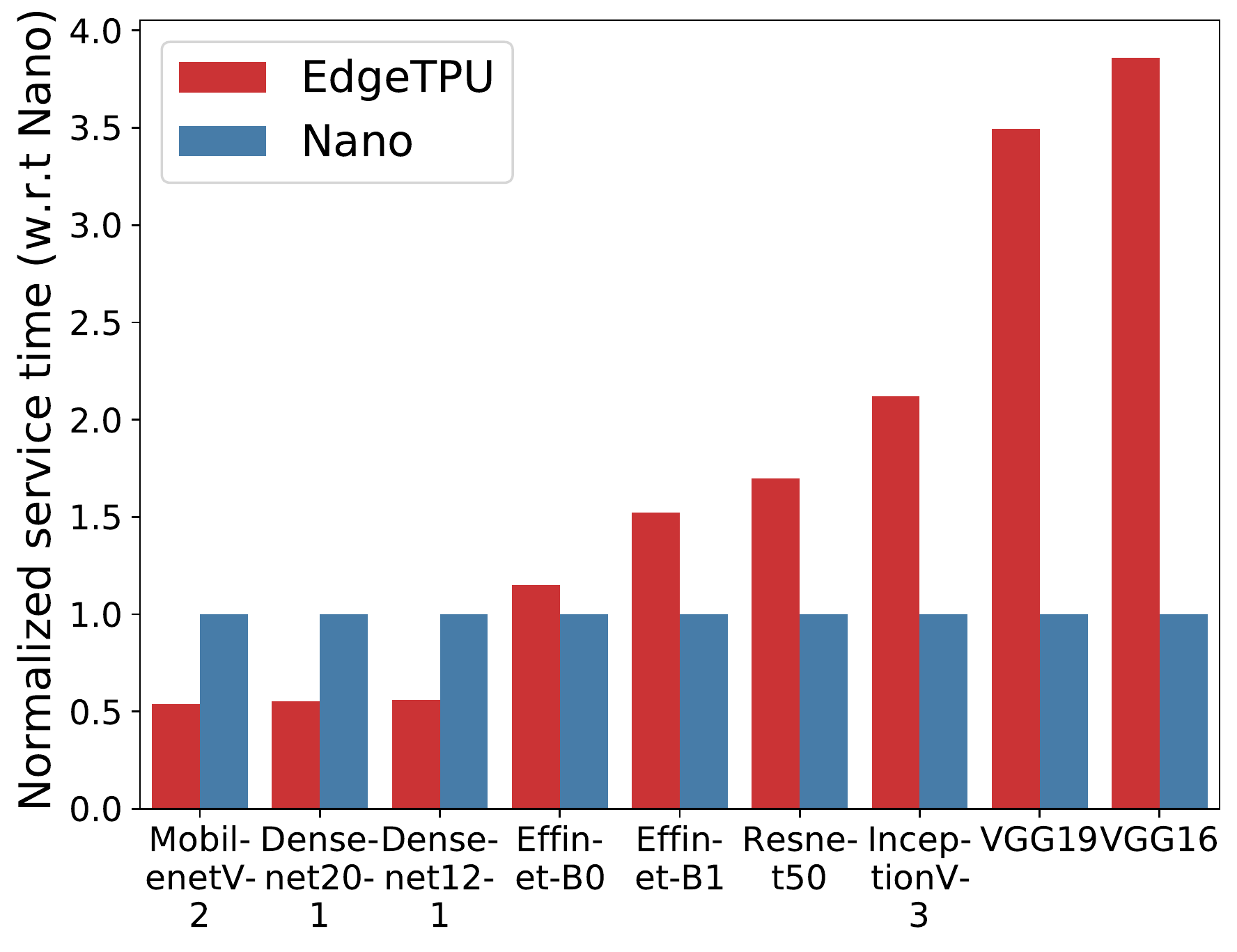}
\captionsetup{belowskip=0pt, aboveskip=5pt}
\caption{Normalized service time for various DNN models.}
\label{fig:heterogeneous_profile}%
\end{minipage}
\hspace{0.4in}
\begin{minipage}{0.45\linewidth}
% TPU Validation
    \centering
	\includegraphics[width=\textwidth]{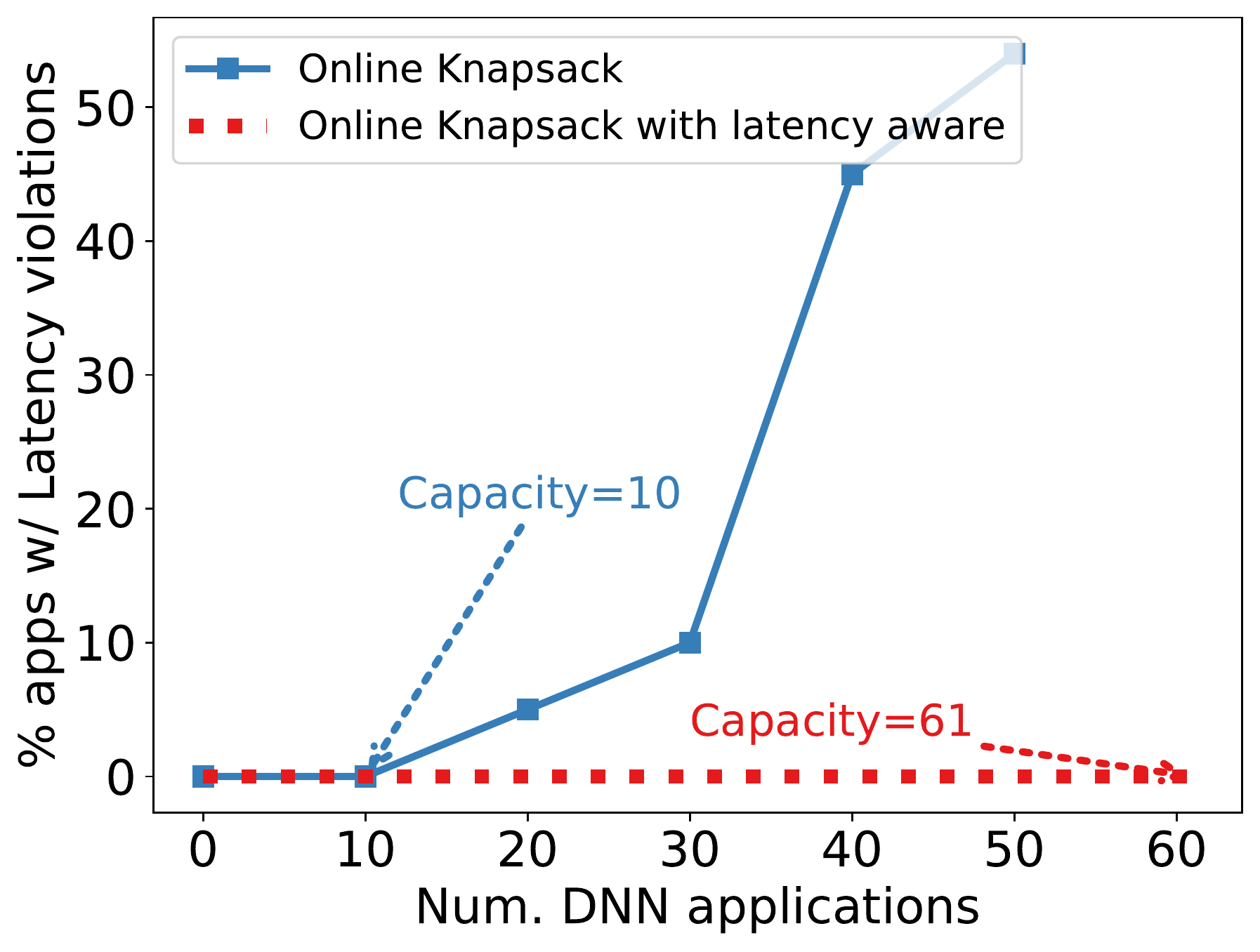}
 \captionsetup{belowskip=0pt, aboveskip=5pt}
	\caption{Latency violations under different co-location methods}
% 	Our latency aware method places 46 models on the 8-node cluster with zero SLO violation.}
	\label{fig:edgetpu-slo}
\end{minipage}
\end{figure*}

We start with profiling of various DNNs to measure their service times in isolation on a  Nano GPU and an edgeTPU.
 Figure \ref{fig:heterogeneous_profile} shows the normalized execution time of various models. Interestingly, the figure shows that smaller DNN models run faster on the TPU than the GPU, while larger ones run faster on the GPU than the TPU.
This is because the edgeTPU is an ASIC designed specifically for DNN inference and has higher operations per second than more general-purpose GPU device. But it has a small (i.e., 8 MB) device or on-chip memory. To run models with memory footprint larger than this limit, it employs host, or off-chip memory, to store the extra model parameters. This incurs a context switch overhead since accessing off-chip memory is much slower than accessing on-chip memory, and hence, larger DNNs  have worse performance than smaller DNNs on EdgeTPU. On the other hand, GPU does not have this problem as it can store all DNN runtime and parameters in its larger device memory; the reduction in context switch overhead offsets the  somewhat higher execution times, yielding lower service times for larger models.

%we would prefer to place those TPU-unfriendly DNNs on GPU. Thus, to achieve maximal sharing, a heterogeneity-aware policy is needed.  Moreover, DNN on-chip memory size also affects the switch overhead on EdgeTPU. When performing context switch, EdgeTPU needs to load the on-chip part of the incoming model into the device memory. As a result, the context switch overhead has a linear relationship with DNN on-chip memory size as shown in figure \ref{fig:tpu_switch_overhead}.  

Our  results show that neither the TPU nor GPU is optimal for all DNN models, and  the optimal choice will depend on the characteristics of each  model. This result also motivates the need for our heterogeneous placement technique discussed in Sec \ref{sec:placement_enhacement}.

\subsubsection{Efficacy of our placement algorithm} 
We conduct an experiment to show the efficacy of our online knapsack placement with latency constraints. We do so by comparing our approach with  a traditional  online knapsack placement approach that is used for cloud application placement.  As noted earlier,  traditional knapsack placement is latency oblivious and performs placement using additive resource packing.  We construct application arrival traces with increasing number of application and place them on our edge cluster using the two placement methods and measure whether there are any response time violations when subjecting applications to the specified request rates.
Figure~\ref{fig:edgetpu-slo} show our results. As shown, the latency-oblivious traditional knapsack see latency violations even with a small number of co-located applications, and the number of applications seeing violations increases with more application arrivals. When we try to load more than 50 application with latency-oblivious policy, the system crashes. In contrast, our latency-aware policy is able to place 61 applications onto the cluster, with no latency violations for any application. 
%This yields a 6-fold improvement in cluster capacity using our approach.

Next we run a simulation experiment where vary the number of applications that need to be placed from 10 to 70, and construct 1000 random arrival traces for each data point. We try to place applications in each trace on a ten node cluster of discrete GPUs using both knapsack methods and determine whether each placement is successful (i.e., whether the applications fit on the cluster and if there are any latency violations as per our queuing equations).  Fig \ref{fig:sucess_rate} shows the fraction of workloads that can be successfully placed on the cluster using online knapsack placement and our latency-aware method. As the number of applications increase,  the probability of successful placement for a random arrival pattern decreases for both methods. However the success rate decreases much faster for the latency-oblivious traditional knapsack due the higher rate of response time violations, while our method can sees higher probability of successful placement.
For a 90\% or higher cutoff success rate, the traditional method can place at most 30 applications, while our method can place nearly 70 applications, yielding a 2.3X increase in cluster capacity.

%it is less likely the online Knapsack method can place the workload successfully without SLO violation. But our latency-aware method can still place the workload successfully with high probability even when the workload size is large. 

\subsubsection{Efficacy of our heterogeneous placement} 
\begin{figure*}
\centering
\captionsetup{justification=centering}
\begin{minipage}{0.45\linewidth}
% Single Model Nano Validation
    \centering
	\includegraphics[width=\textwidth, valign=t]{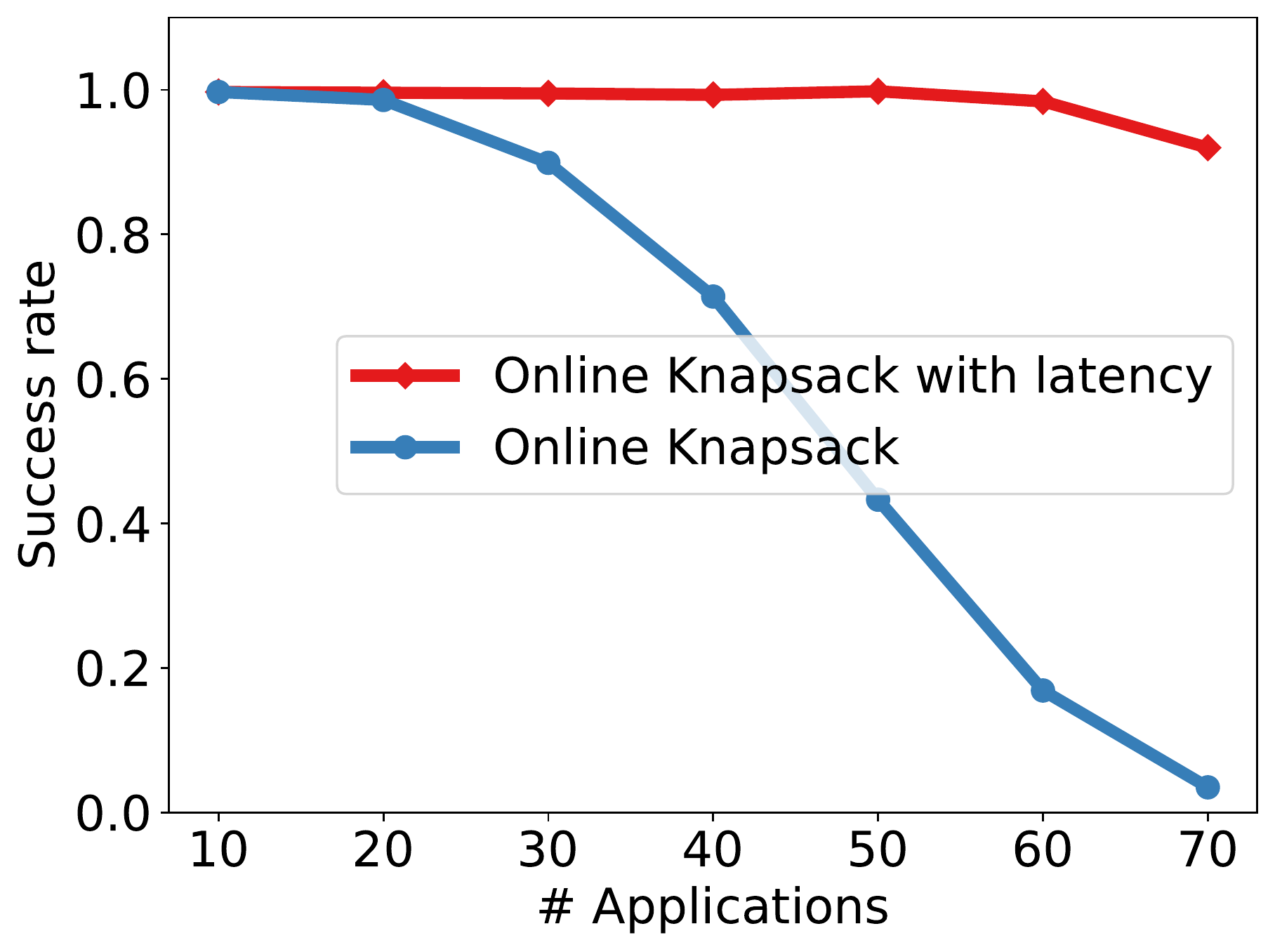}
	\captionsetup{belowskip=0pt, aboveskip=5pt}
	\caption{Latency aware vs traditional knapsack}
	\label{fig:sucess_rate}
\end{minipage}
\hspace{0.4in}
\begin{minipage}{0.45\linewidth}
% TPU Validation
    \centering
	\includegraphics[width=\textwidth,valign=t]{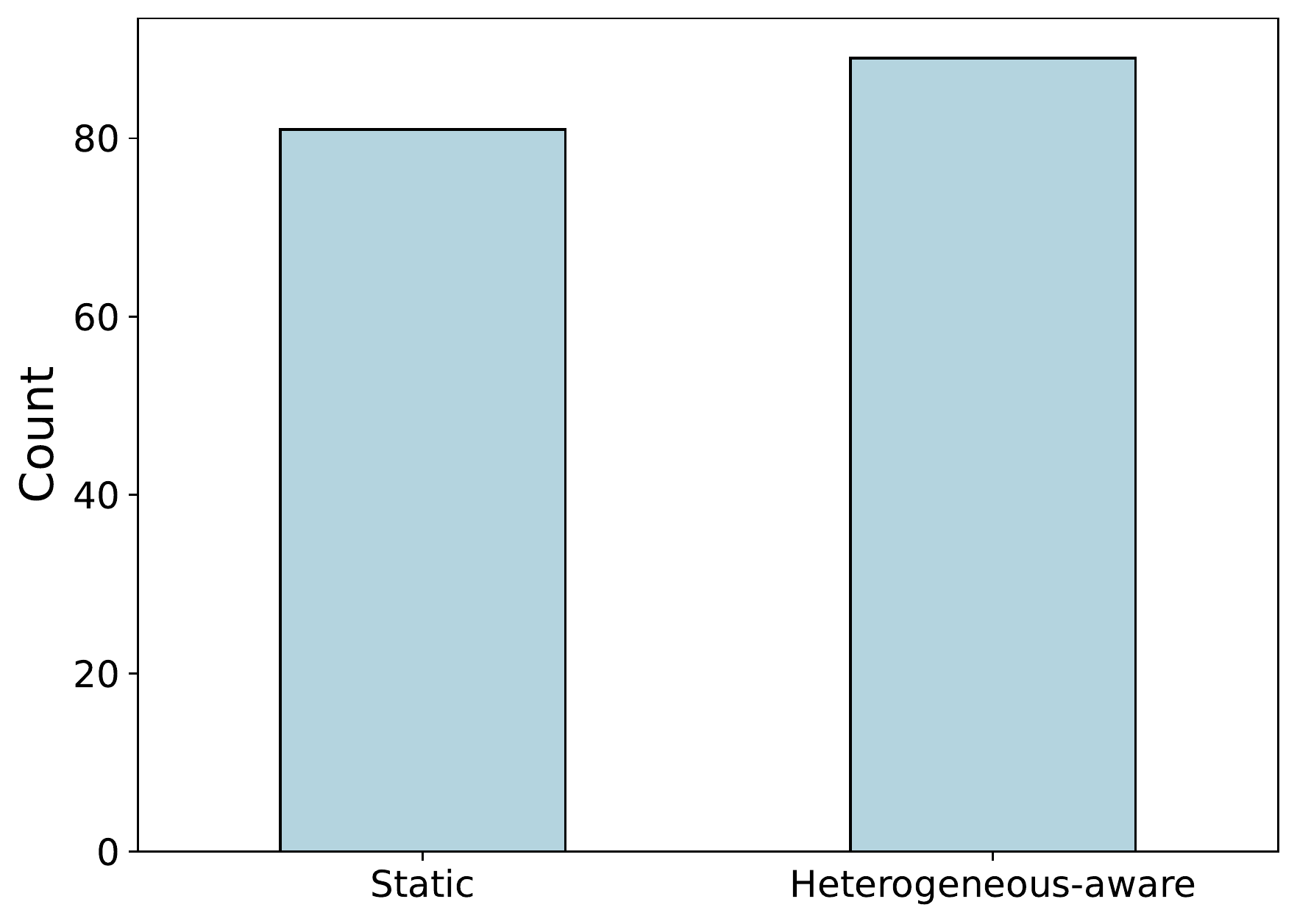}
	\captionsetup{belowskip=0pt, aboveskip=5pt}
	\caption{Benefits of heterogeneous placement.}
% 	Our latency aware method places 46 models on the 8-node cluster with zero SLO violation.}
	\label{fig:heter_vs_random}
\end{minipage}
\end{figure*}

We next show the efficacy of our heterogeneous placement enhancement. We conduct an experiment where we construct an arrival sequence where each application randomly makes a static choice of  a GPU or TPU accelerator and we use our latency-aware online knapsack to make placement decisions. We compare this static placement policy with our heterogeneous placement approach where the choice of the accelerator is made dynamically by our placement algorithm. Figure \ref{fig:heter_vs_random} shows the number of applications placed onto the cluster in each case. As can be seen, when a static choice is made, some applications can make a bad choice and use more resource than they need (e.g. as see in fig \ref{fig:heterogeneous_profile} VGG16 has 4$\times$ more service time on TPU than on GPU). In case of heterogeneous placement, the best accelerator is chosen for each application and resources are effectively used to pack more application. The heterogeneous policy yield a cluster capacity improvement of 10\%. This experiment shows that Ibis is able to maximize resource sharing while maintaining response time guarantees.

%We first conduct an experiment to co-locate a mix of IoT applications on  the cluster using AI-as-a-Service. In this experiment, we compare three policies :1) a latency-oblivious policy that only looks at memory footprint for placement (i.e. using a knapsack approach); 2) our latency-aware grouped placement policy that select feasible nodes but randomly choose device type; 3) our latency \& heterogeneity-aware grouped placement policy that selects the optimal accelerator type that offers lowest service time. Figure~\ref{fig:edgetpu-slo} depicts our results. As shown, the latency-oblivious policy see latency violations even with a small number of co-located applications, and the number of applications seeing violation increase with more customer arrivals. When we try to load more than 50 customers with latency-oblivious policy, the system crashes. In contrast, our latency-aware policy is able to place 61 customers on to the cluster, with no latency violation for any of the customers. Furthermore, our latency \& heterogeneity-aware policy is able to place 70 customers on to the cluster without any latency violations. These validate that Ibis is able to maximize resource sharing while maintaining response time guarantees.

\subsubsection{Efficacy of AIaaS Placement} Next, we conduct an experiment to compare the degree of sharing achieved using   AI-as-a-Service applications, user-trained models, and a mix of both.  We construct a trace of application arrivals and use our algorithms to place these applications using all three approaches.
Figure \ref{fig:placement_user_iaas} shows the number of applications that can be placed on the cluster in each case. Since AI-as-a-Service applications share a DNN model, they can achieve a greater degree of resource sharing. The figure shows that each Jetson Nano node can hold only 2 to 3 DNN models since TensorRT run time takes around 800 MB aside from the models' actual memory requirements. %(see table \ref{tab:all-models}).
However, in the AI-as-a-Service, the system can host double the number of customers since  sharing amortizes the memory footprint of the TensorRT runtime. 
%Customers need only a small frontend container due to backend sharing, allowing for better GPU utilization by hosting more customers. 
On the other hand,  in the user-trained case, the memory limit on a node is reduced once 2-3 models are placed, and additional DNN models cannot be placed even though the GPU is underutilized. Finally, the mixed workload achieves co-location performance that lies between the two.

\subsubsection{Evaluation of hotspot migration with token bucket} 
\begin{figure*}[t]
\centering
\captionsetup{justification=centering}
\begin{minipage}{0.43\linewidth}
% TPU Validation
    \centering
	\includegraphics[width=\textwidth]{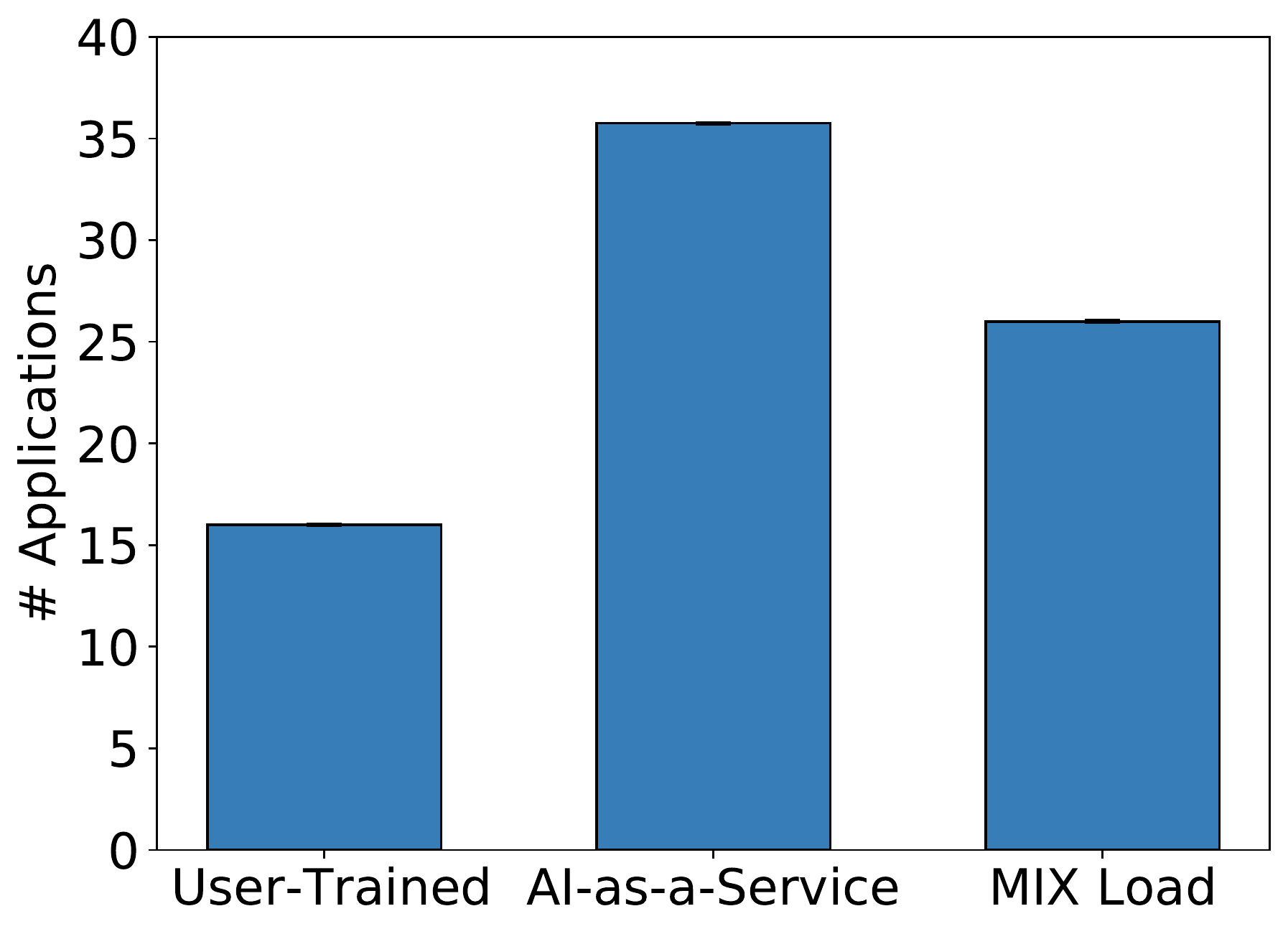}
		\captionsetup{belowskip=0pt, aboveskip=2pt}
	\caption{Max application capacity for AIaaS and user-trained models}
	\label{fig:placement_user_iaas}
\end{minipage}
%\begin{minipage}{0.31\linewidth}
% TPU Validation
%    \centering
%   \includegraphics[width=\textwidth]{imgs/placement_nano_SLA.pdf}
%	\caption{\textcolor{red}{remove}User-Trained models SLO violation rate on Jetson Nano.}
%	\label{fig:placement_nano_sla}
%\end{minipage}
\hspace{0.4in}
\begin{minipage}{0.43\linewidth}
% Single Model Nano Validation
    \centering
	\includegraphics[width=\textwidth]{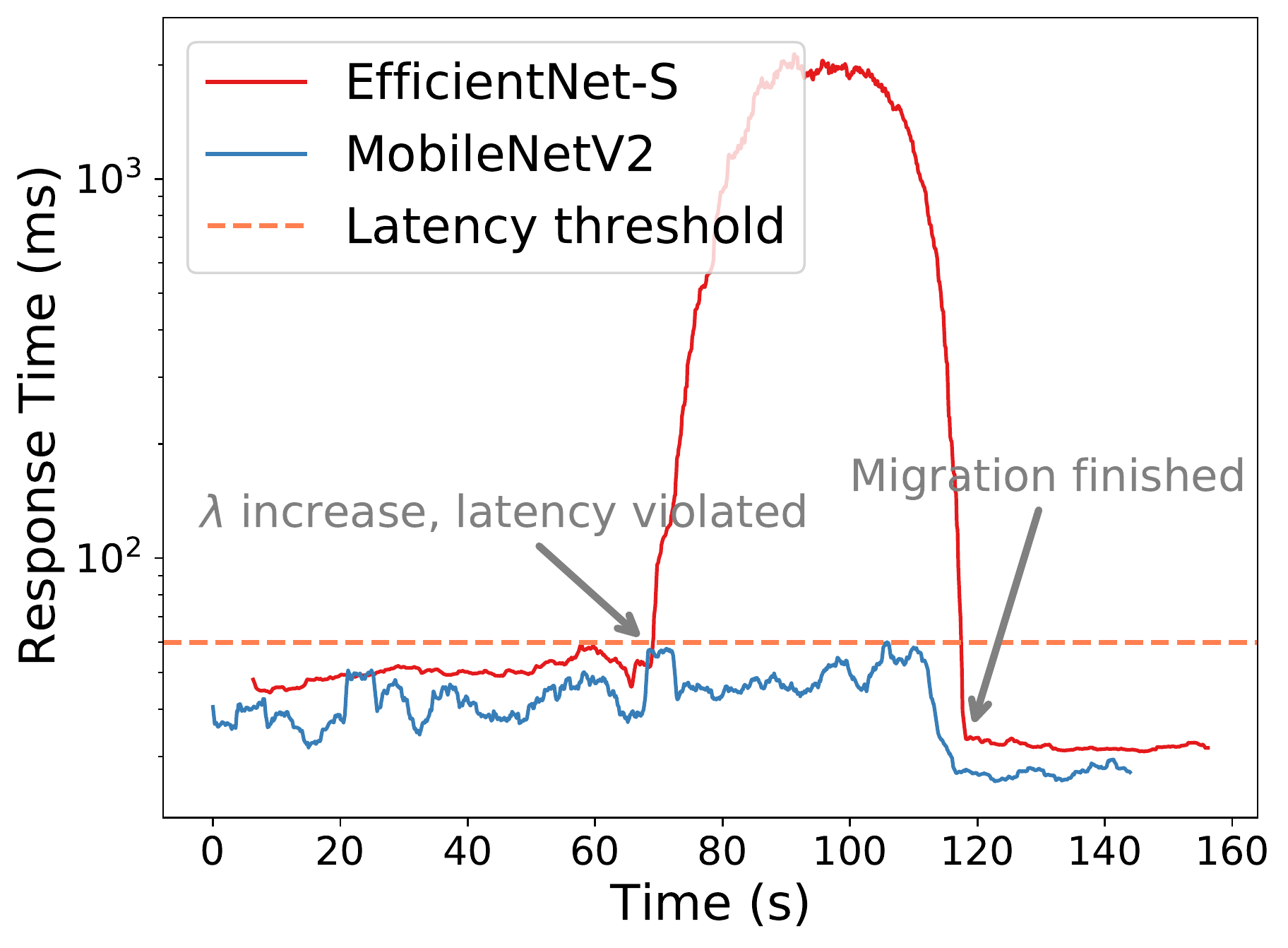}
		\captionsetup{belowskip=0pt, aboveskip=2pt}
	\caption{Hotspot mitigation with token bucket.}
	\label{fig:migration}
\end{minipage}
\end{figure*}

Finally, we conduct an experiment to demonstrate the efficacy of hotspot mitigation with token bucket. We start the experiment with the system moderately loaded with two DNN models --- EfficientNet-S and MobileNetV2. We then increase the request rate of EfficientNet-S beyond the placement time estimate. As shown in figure \ref{fig:migration}, the workload increase causes the response time of EfficientNet-S to exceed the latency threshold but does not affect the MobileNetV2 because of the token bucket. Our system monitors the response time seen by customers and triggers a migration when violations are observed. The overloaded application is migrated to a new node with sufficient idle capacity, causing the response time of the overloaded application to fall below the latency violation threshold. This experiment shows the ability of our techniques to mitigate hotspots dynamically.

% Tiered Edge Cloud Architecture
%\input{src/architecture}

% Placement
%\input{src/placement}

% Experiment
%\input{src/experiment.tex}

\section{Related Work}

\noindent\textbf{Edge inference.} The promise of edge computing is to provide lower latency for real-time applications by eliminating the latency of offloading to remote clouds~\cite{salehe2019videopipe, Satyanarayanan2017,zhang2017towards, AI2018}. Hardware accelerators such as GPUs and TPUs provide a powerful add-on for edge infrastructure \cite{Chen2019,Zhou2019, AppleDetection}. Many researchers have studied and characterized the performance of various edge accelerators~\cite{hadidi2019characterizing,qianlin2020}. Wu et al.~\cite{wu2019fpga} demonstrate  an FPGA-accelerated and general-purpose distributed stream processing system for Edge stream processing.

\noindent \textbf{Managing inference workloads.} Building SLO aware inference systems have been widely discussed \cite{Nexus, MArk, soifer2019deep}. Zhang et al.~\cite{MArk} proposed a scalable system for DNN inference workloads, where different classes of resources are provisioned to allow efficient and effective scaling. Similarly, Nexus~\cite{Nexus} provides a solution for deploying multi-level AI workload utilizing batching to decrease the processing time. Soifer et al.~\cite{soifer2019deep} provide insights on the inference infrastructure at Microsoft, which duplicates requests with cross cancellation tokens to ensure predictable response times. In our work, we provide a proactive estimation tool for predicting the performance of DNNs and provide an latency and heterogeneity aware scheduling mechanism for edge deployments.

%\noindent \textbf{Multi-tenancy on accelerators.} Due to resource scarcity, edge resources are usually shared across multiple users, with many recent works discussing the intricacies of resource sharing for GPUs~\cite{Amert2017, Yang2018, PERSEUS}. Perseus~\cite{PERSEUS} studies the performance effects of running multiple DNNs on shared GPUs. On the other hand, we provide a theoretical foundation that captures these effects in discrete and edger embedded GPU. However, as far as we know we, are the first to discuss the effect of resource sharing for TPUs. 

\noindent \textbf{Multi-tenancy on accelerators.} Sharing of GPUs across applications has been studied for cloud servers \cite{gslice, olympian, gpushare}. Olympian \cite{olympian} and GPUShare \cite{gpushare}  focus on sharing a single GPU across multiple users, while GSLICE \cite{gslice} focuses cluster-level sharing. %These papers try to minimize the effect of resource sharing, however, we provide a framework for understanding and predicting the effect of this multiplexing.
In contrast to these efforts, we focus on analytic models, using queueing theory, to enable GPU or TPU multiplexing while providing response time guarantees.

\noindent \textbf{Model-aware placement.} Model (queueing models) have been used extensively to monitor and predict the performance of traditional processing units (i.e., CPU) \cite{Urgaonkar2005, cloudscale, Grandhi2019Queueing, HarcholBalter2005MultiServerQS, serverFarmsHarchol, waiting-game}.  There has been recent work on capturing performance of accelerators using queueing models. The work in \cite{GPU_queue_model} presents a queueing model to captures effect of GPU batching of a single application. 
We focus on the effect of multi-tenancy and cover multiple accelerator architectures. 

% Conclusion
\section{Conclusion}

In this paper, we presented analytic models to estimate the latency behavior of DNN inference workloads on shared edge accelerators, such as GPU and edgeTPU, under different multiplexing and concurrency behaviors.  We then used these  models to design resource management algorithms to intelligently co-locate multiple applications onto edge accelerators while respecting their latency constraints.   Our results showed that our models can accurately predict the latency behavior of DNN applications on shared nodes and accelerators, while  our  algorithms  improve resource sharing by up to 2.3X when providing response time guarantees.

%%
%% The acknowledgments section is defined using the "acks" environment
%% (and NOT an unnumbered section). This ensures the proper
%% identification of the section in the article metadata, and the
%% consistent spelling of the heading.
%\begin{acks}
%To Robert, for the bagels and explaining CMYK and color spaces.
%\end{acks}

%%
%% The next two lines define the bibliography style to be used, and
%% the bibliography file.
\bibliographystyle{ACM-Reference-Format}
\bibliography{references}

%%
%% If your work has an appendix, this is the place to put it.
%\input{src/appendix}

\end{document}